\documentclass[a4paper,fleqn,final]{cas-dc}

\usepackage[numbers]{natbib}
\usepackage{subcaption}

\setcitestyle{numbers,square}

\usepackage{graphicx}
\usepackage{array}
\usepackage{caption}
\usepackage{stfloats}
\usepackage{booktabs} 
\usepackage{float}

\def\tsc#1{\csdef{#1}{\textsc{\lowercase{#1}}\xspace}}
\tsc{WGM}
\tsc{QE}
\tsc{EP}
\tsc{PMS}
\tsc{BEC}
\tsc{DE}

\graphicspath{{Figure/}}

\begin{document}

\let\WriteBookmarks\relax
\def\floatpagepagefraction{1}
\def\textpagefraction{.001}

\shorttitle{Game transition based on reputation}

\shortauthors{Hongyu Yue et~al.}


\title {Dynamic Evolution of Cooperation Based on Adaptive Reputation Threshold and Game Transition}

\author[1]{Hongyu Yue}[style=Chinese]

\author[1]{Xiaojin Xiong}[style=chinese]

\author[1]{Minyu Feng}[style=chinese,orcid=0000-0001-6772-3017]
\cormark[1]
\ead{myfeng@swu.edu.cn}

\author[2]{Attila Szolnoki}

\affiliation[1]{organization={College of Artificial Intelligence},
    addressline={Southwest University}, 
    city={Chongqing},
    postcode={400715}, 
    country={PR China}}

\affiliation[2]{organization={Institute of Technical Physics and Materials Science},
    addressline={Centre for Energy Research}, 
    city={Budapest},
    postcode={H-1525},
    country={Hungary}}

\begin{abstract}
In real-world social systems, individual interactions are frequently shaped by reputation, which not only influences partner selection but also affects the nature and benefits of the interactions themselves. We propose a heterogeneous game transition model that incorporates a reputation-based dynamic threshold mechanism to investigate how reputation regulates game evolution. In our framework, individuals determine the type of game they engage in according to their own and their neighbors'  reputation levels. In turn, the outcomes of these interactions modify their reputations, thereby driving the adaptation and evolution of future strategies in a feedback-informed manner. Through simulations on two representative topological structures, square lattice and small-world networks, we find that network topology exerts a profound influence on the evolutionary dynamics. Due to its localized connection characteristics, the square lattice network fosters the long-term coexistence of competing strategies. In contrast, the small-world network is more susceptible to changes in system parameters due to the efficiency of information dissemination and the sensitivity of strategy evolution. Additionally, the reputation mechanism is significant in promoting the formation of a dominant state of cooperation, especially in contexts of high sensitivity to reputation. Although the initial distribution of reputation influences the early stage of the evolutionary path, it has little effect on the final steady state of the system. Hence, we can conclude that the ultimate steady state of evolution is primarily determined by the reputation mechanism and the network structure.
\end{abstract}


\maketitle

\section{Introduction}
In the study of evolutionary game theory, understanding the mechanism of cooperation and its maintenance conditions has been one of the main issues that have attracted widespread attention in the academic community~\cite{weibull1997evolutionary}. Individuals often need to sacrifice their interests to achieve the general interests of the group. Choosing cooperation seems to violate the basic principle of ``survival of the fittest'' in Darwin's natural selection~\cite{axelrod1981evolution}. However, in the real world, whether it is a microscopic biological system or a macroscopic human society, the phenomenon of cooperation is always ubiquitous and plays an irreplaceable role in the promotion of social progress and the evolution of civilization~\cite{perc2017statistical,ostrom1990governing}. As an interdisciplinary theoretical framework that connects biology, economics, and social sciences, evolutionary game theory was proposed to explain such phenomena and theoretically provides a solid analytical tool and model framework for the formation and stability of cooperation~\cite{smith1973logic}. In recent years, it has also been widely applied to the analysis of cooperation in digital economic systems and data governance~\cite{li2023open}. In this framework, typical models such as the prisoner's dilemma~\cite{poundstone2011prisoner,yang2024interaction,yue2025coevolution}, the snowdrift game\cite{doebeli2005models,ding2024emergence}, and the stag hunt game~\cite{luo2021evolutionary,belloc2019intuition,battalio2001optimization} are widely used to describe the conflict between cooperation and defection in the real world and their evolutionary path, becoming an important basis for exploring the mechanism of cooperation.

In the framework of traditional evolutionary game theory, the introduction of spatial structure is considered as one of the important mechanisms to promote the evolution of cooperation\cite{szabo_pr07,wang_z_epjb15}. In an early work, Nowak and May\cite{nowak1992evolutionary} pointed out that although the defection strategy has a significant evolutionary advantage, the clustering effect caused by local interactions can effectively promote the formation and maintenance of cooperation. This discovery has greatly promoted academic research on game behavior in complex networks, and related results generally believe that spatial structure plays a positive role in most social dilemmas~\cite{van2013spatial, roca2009effect}. Zeng {\it et al.} proposed a complex network model with power-law activating patterns, revealing how heterogeneous interaction frequencies affect the coevolution of structure and behavior~\cite{zeng2025complex}. However, subsequent studies have also found that spatial structure does not promote cooperation in all types of games. For example, in the snowdrift game, spatial structure may inhibit the spread of cooperation if imitation is applied~\cite{hauert2004spatial}. This observation warns us that the impact of the network topology on the evolution of cooperation is more subtle and may depend on the specific setting of game types and the interaction rules.

Over the years, scholars gradually realized that static network structures cannot fully capture the dynamic and heterogeneous nature of individual interactions observed in real societies~\cite{pacheco2007evolution,miritello2013temporal,su2023strategy}. Feng {\it et al.} investigated the information dynamics in evolving networks through the lens of the birth-death process, and highlighted how the interplay of random drift and natural selection shapes the evolution of individual states and interaction structures~\cite{feng2024information}. These findings exemplify a broader shift in evolutionary game research from fixed interaction frameworks to adaptive systems in which both strategies and network topologies evolve. In recent years, coevolutionary game theory has gradually emerged~\cite{perc2010coevolutionary}, coupling the evolution of individual strategies with the evolution of interaction structures~\cite{xiong_xj_c24,wang2013interdependent,wang_q_csf25}, and has become a new paradigm for understanding the evolution of cooperation. Individuals not only adjust their strategies but also actively change their interactions based on benefit feedback, such as disconnecting from defectors and choosing new partners~\cite{pacheco2006coevolution,szolnoki_njp09}, thereby dynamically reconstructing their local networks\cite{wang2014self}. The type of co-evolutionary mechanism~\cite{stewart2014collapse}, such as link reciprocity and structural adaptability, has not only been theoretically proven to improve the level of cooperation significantly but has also been widely supported by human experiments~\cite{efferson_s08,dong_z_srep18}. More recently, the focus has extended beyond structural dynamics to include the evolution of preference frameworks themselves, transitioning from outcome-based models to language-based and norm-sensitive approaches~\cite{capraro2024outcome}, thereby enriching the theoretical foundations for understanding social decision making.

It is worth noting that the study of coevolutionary games has gradually expanded from focusing solely on structural evolution to incorporating heterogeneity in individual attributes. Among these, mechanisms centered on reputation have attracted increasing scholarly attention~\cite{zhang2018coevolution,wang2017inferring}. Reputation~\cite{wang2012inferring, gallarta2024emergence}, as a cumulative reflection of an individual’s behavioral history, influences not only the dissemination of strategies but also the selection of interaction partners~\cite{nowak1998evolution}, thereby shaping a dynamic and feedback-sensitive game environment. These developments offer novel theoretical tools and simulation frameworks for capturing the complex evolutionary pathways of cooperation in real-world societies. In their review, Xia {\it et al.} comprehensively illustrated the pivotal role of reputation and reciprocity in the evolution of cooperation~\cite{xia2023reputation}. They emphasized that reputation, as a record of social conduct, governs strategic decisions while simultaneously reinforcing the stability and propagation of cooperation when interlinked with direct and indirect reciprocity mechanisms. Further advancing the line of inquiry, Hu {\it et al.} introduced an adaptive reputation model demonstrating that, in dynamic social networks, individuals adjust their trust levels in real-time based on interactive experiences, thereby fostering more resilient cooperation structures~\cite{hu2024reputation}. The adaptive process significantly enhances both the level of trust and the prevalence of cooperation across the network. On a more theoretical level, Capraro and Perk provided a formal foundation for modeling moral preferences, analyzing how social norms, reputation, and individual decisions coevolve~\cite{capraro2021mathematical}.

In real-world social systems, interactions between individuals do not take place within a uniform or static game environment. Instead, they exhibit a high degree of heterogeneity and dynamism. Su {\it et al.} systematically proposed the concept of game transitions, suggesting that individuals may encounter different types of strategic interactions over time, even within a single evolutionary framework~\cite{su2019evolutionary}. The approach fundamentally extends traditional models by allowing the payoff environment to evolve alongside strategy dynamics. Building upon this foundation, Feng {\it et al.} further incorporated Markov processes to govern the transitions between games, providing a stochastic framework that captures the probabilistic and temporal characteristics of such environmental transitions~\cite{feng2023evolutionary}. The mechanism by which games are transformed can be driven by various factors, among which reputation serves as a crucial social signal that influences access to cooperation opportunities and the overall quality of interactions~\cite{feng2024evolutionary,zhang2025evolution,li2021reputation}. In general, individuals with higher reputations tend to obtain more valuable cooperation resources and are more likely to engage in high-payoff games, while those with lower reputations are often marginalized and engage in low-payoff or even exclusionary interactions. Building on these observations, we propose a heterogeneous game transition model that incorporates a dynamic reputation threshold mechanism. In our model, individuals are dynamically classified as either high- or low-reputation agents based on the average reputation level of the population, which in turn determines the type of game they engage in either a high-value or low-value game. Specifically, high-reputation individuals participate in high-value games, while low-reputation individuals participate in low-value games. When individuals of different reputation categories interact, game transitions occur with probability. Furthermore, individual reputations are updated in response to their strategies, and the feedback mechanism subsequently influences their strategic preferences and interaction patterns in future rounds, forming a feedback-driven evolutionary dynamic system.

Our paper is structured as follows. We first describe our model in Section~\ref{sec: model}. It is a game transition model based on adaptive reputation thresholds and a policy update rule based on fitness. In Section~\ref{sec: results} we present in detail the experimental results obtained in a wide variety of parameter values and different network structures. Finally, in Section~\ref{sec: conclusion} we summarize our research results and sketch some potential future developments.

\section{Model}
\label{sec: model}

In interpersonal actions, the reputation of an agent plays a crucial role in determining the quality of the interaction. An individual's social evaluation not only affects his or her position in the group but also directly determines how he or she interacts with others. In this section, we propose a model for the evolution of the game transition based on reputation feedback to characterize how individuals choose game objects and game types based on their reputation values on social networks. The model assumes that each agent is represented as a node in the network and that the edges between nodes represent possible interactions between agents. Each agent has a dynamically changing reputation value and interacts with other neighbors randomly at each time step, and the specific game-type involved is determined by the reputation of the two interacting parties.

\begin{figure*}[htbp]
    \centering
    \includegraphics[width=\linewidth]{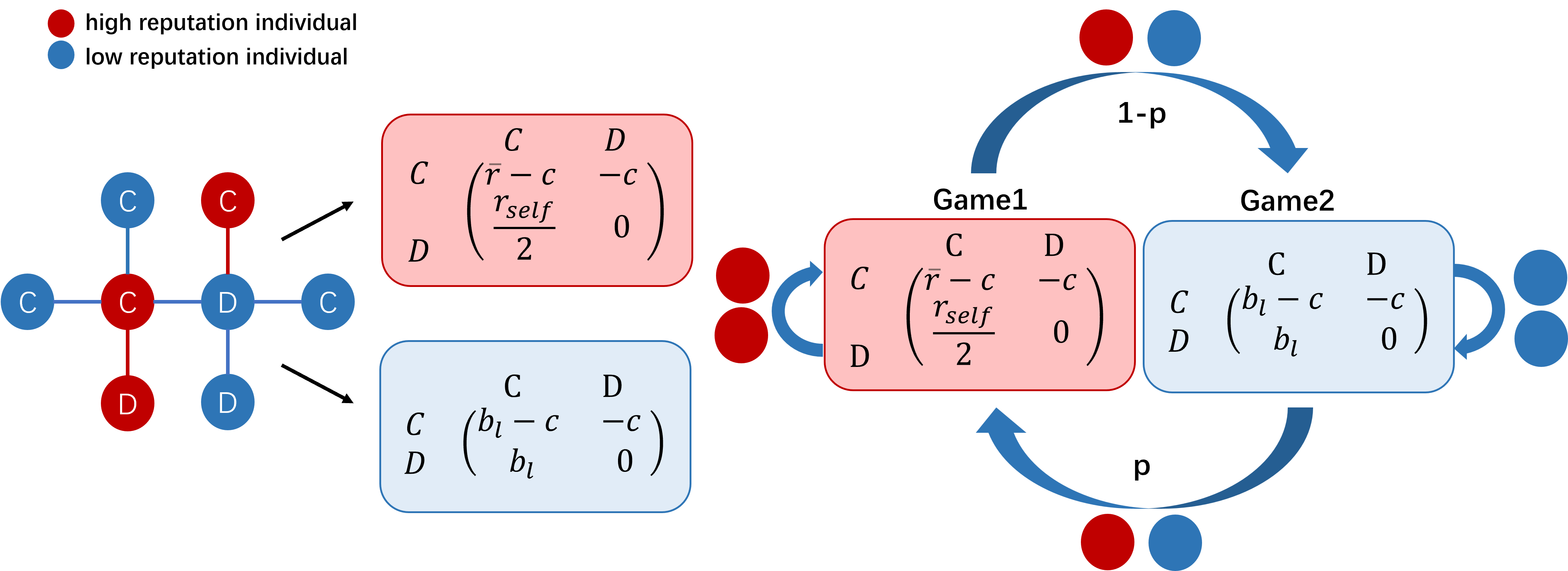}
    \caption{\footnotesize \textbf{Reputation-based game type transition mechanism.} The figure illustrates how the actual social game type between individuals is affected by their reputation status. The model introduces an adaptive reputation threshold to dynamically divide agents into high-reputation individuals (red) and low-reputation individuals (blue). A high-value game is played when both interacting individuals have high reputations, whereas a low-value game occurs when both have low reputations. In interactions between a high-reputation and a low-reputation individual, the game type alternates probabilistically between these two. For specific rules and parameter settings, please refer to the main text.}
    \label{fig1}
\end{figure*}

\subsection{Game Classification}

In the evolutionary process, agents may participate in different types of games, where each game requires individuals to choose between two fundamental strategies: cooperation (\(C\)) or defection (\(D\)). This decision determines the final payoff, which depends on the specific combination of their chosen strategies. In a given game \(l\), two agents make their decisions simultaneously and independently. If both agents choose to cooperate (\(C, C\)), each will receive a payoff \(R_l\) and if both defect (\(D, D\)), each receives a lower payoff \(P_l\). When one defects and the other cooperates, the defector earns a higher payoff \(T_l\), while the cooperator receives a lower payoff \(S_l\).

As a baseline, when the agents’ reputations are low, the interaction follows a classic Prisoner's Dilemma structure, in which the payoff values satisfy the following condition:
\[
T_l > R_l > P_l > S_l.
\]
\par In this scenario, we assume that cooperation requires a cost \(c\), but brings potential benefits \(b_l\), which may vary with the context of the interaction. The corresponding payoff matrix is:
\[
\begin{bmatrix}
b_l - c & -c \\
b_l & 0
\end{bmatrix},
\]
where the classic structure of the Prisoner's Dilemma is retained: mutual cooperation provides a benefit \(b_i\), while defection yields a higher payoff to the defector at the expense of the cooperator. It captures the fundamental social dilemma presented in many real-world interactions, where cooperation is costly but beneficial in the long term, and defection offers an immediate but often unsustainable advantage.

However, in many real-world social and economic systems, interactions between agents are significantly influenced by reputation. Reputation serves as a critical social signal, reflecting an agent’s history of cooperation, social standing, or perceived trustworthiness. Agents with higher reputations are more likely to be trusted, invited into valuable collaborations, and given access to superior resources. In this sense, reputation functions as a form of social capital, translating into tangible strategic advantages.

To capture the social dynamic, we extend the baseline model by introducing a heterogeneous, reputation-driven game. In the modified framework, an agent’s reputation influences the payoffs they receive in interactions. Specifically, when an agent’s reputation exceeds a given threshold, they gain access to high-value interactions, in which the rewards from cooperation increase with the reputations of both participants. The latter element reflects empirical observations that high-reputation individuals tend to engage in more beneficial and sustained cooperation. The payoff matrix for such interactions is given by:
\[
\begin{bmatrix}
\bar{r} - c & -c \\
\frac{r_{\text{self}}}{2} & 0
\end{bmatrix},
\]
where \(\bar{r}\) denotes the average reputation of the interacting agents \(i\) and \(j\), reflecting their joint social standing, and \(r_{\text{self}}\) represents the reputation of the agent in question. The parameter \(c\) indicates the cost of cooperation. In order to highlight the role of reputation as a factor promoting mutual trust, we assume that when two agents choose to cooperate, the cost of cooperation is shared equally by both parties, thereby effectively reducing individual burdens and promoting cooperation.

By embedding reputation directly into the payoff structure, we capture the reputation-mediated enhancement of cooperation payoffs. Agents with higher reputations are more likely to access superior interaction opportunities and obtain greater benefits because reputation serves as a form of social capital that influences both the likelihood of being chosen as a partner and the expected payoff from interactions. This assumption is consistent with empirical observations: high-reputation agents tend to form more stable, productive, and mutually beneficial relationships, supported by stronger mutual trust.

In contrast, interactions involving low-reputation agents often lack trust and are susceptible to opportunistic behavior. Such interactions are characterized by a lower willingness to cooperate, a higher risk of cooperation failure, and subsequently diminished payoffs. This reputation asymmetry highlights the critical role of social trust structures in dynamic strategic environments and illustrates how historical behavior embodied in reputation can have profound effects on future strategies.

\subsection{Reputation-Based Game Transition}

In real-world social systems, an individual's reputation plays a pivotal role in shaping their interaction patterns and the benefits they can attain. Individuals with higher reputations are typically more likely to be trusted and included in profitable cooperative endeavors, whereas those with lower reputations may be marginalized and confined to lower-payoff interactions. Such dynamics underscore the stratified nature of social cooperation, where reputation serves as an informal credential for access to high-value opportunities.

To capture this element in our model, we introduce an \textit{adaptive reputation threshold} \( \theta(t) \), which dynamically separates high-reputation individuals from low-reputation ones. The threshold is designed to evolve with the population's overall reputation level and is determined by the average reputation at time \( t \), expressed as:
\begin{equation}
\theta(t) = \text{avr}_r(t).
\end{equation}

Based on the adaptive threshold, each individual \( i \) at time \( t \) can be categorized as follows:
\begin{equation}
i \in
\begin{cases}
H, & r_i(t) > \theta(t) \quad \text{(high-reputation individual)} \\
L, & r_i(t) \leq \theta(t) \quad \text{(low-reputation individual)}
\end{cases}.
\end{equation}

To further refine the interaction structure, we introduce a reputation-driven game selection mechanism that depends on the reputation categories of the interacting pair \( (i, j) \). When both individuals have high reputations, \( (i, j) \in H \times H \), they engage in a \textit{high-value game}, which yields mutually beneficial outcomes with greater potential payoffs. Conversely, when both individuals have low reputations, \( (i, j) \in L \times L \), they participate in a \textit{low-value game}, where cooperation is limited and payoffs are modest. In mixed-reputation pairings, \( (i, j) \in H \times L \cup L \times H \), the game type is determined probabilistically: a high-value game occurs with probability \( p \), and a low-value game with probability \( 1 - p \). The stochastic mechanism captures the asymmetry of real-world interactions. Namely, high-reputation individuals are more likely to access valuable opportunities, while their low-reputation counterparts are not categorically excluded, but face restricted chances to engage in high-payoff interactions. By allowing probabilistic transitions between game types, we can mitigate rigid stratification and foster mobility within the reputation hierarchy.

The reputation-based game transition structure governs both the conditions of participation and the subsequent consequences for the evolution of individual reputation and strategic behavior. These two components are inherently interconnected and co-evolve over time, jointly shaping the system's overall dynamics. In particular, the outcome of each interaction significantly influences the trajectory of an individual's social standing. It is important to note that a player's reputation, being dynamic and context-sensitive, is continuously updated in response to the strategies adopted during pairwise interactions. Individuals are socially rewarded or penalized based on the degree to which their actions conform to the expectation of cooperation. Specifically, cooperation is assumed to enhance an individual's reputation, while defection generally leads to its decline.

The reputation update rule for individual \(i\) at time step \(t\) is defined as:
\begin{equation}
r_i(t) = \left[ r_i(t-1) + \sum_{j \in N_i} \Delta r_{ij} \right]_0^2,
\end{equation}
where $N_i$ represents the set of neighbors of agent i and \(\Delta r_{ij}\) represents the change in reputation after interacting with j, which is determined by the strategy combination of both agents. The rule is specified as follows:
\begin{equation}
\Delta r_{ij} =
\begin{cases}
\delta, & \text{if }S_i = S_j = C \\
2\delta, & \text{if }S_i \neq S_j \text{ and } S_i = C \\
-2\delta, & \text{if }S_i \neq S_j\text{ and } S_i = D \\
-\delta, & \text{if }S_i = S_j = D
\end{cases} ,
\end{equation}
where \(S_i\) denotes the strategy adopted by individual \(i\), and \(\delta\) is a fixed parameter that quantifies the magnitude of reputation change. This rule differentiates between symmetric and asymmetric interactions, with larger reputation adjustments occurring in the latter case when an individual cooperates while the other defects. In other words, behavior deviating from social norms, whether positive or negative, has a more significant impact on an individual's reputation.

To ensure numerical stability and keep the reputation measure interpretable, we employ a truncation operation that confines \(r_i(t)\) to the interval \([0, 2]\), thereby preventing unbounded growth or underflow:
\begin{equation}
[a]^2_0 =
\begin{cases}
0, & \text{if } a < 0 \\
a, & \text{if } 0 \leq a \leq 2 \\
2, & \text{if } a > 2
\end{cases}.
\end{equation}

As reputation accumulates through repeated interactions, it begins to exert an increasingly significant influence on the evolutionary dynamics of strategies. The dual role of reputation, as both an outcome of past actions and a determinant of future strategic success, establishes a feedback loop that shapes the trajectory of social behavior over time.

In particular, the strategy update process is governed by an individual's fitness, which integrates both payoff-related and social dimensions. Based on this, we define a fitness value as a weighted combination of the individual’s accumulated material payoffs and their current reputation level, reflecting the intuition that both material wealth and social recognition are critical in determining the influence of an individual on the population. The fitness of individual \( i \) at time \( t \) is given by:
\begin{equation}
f_i = m \cdot \Pi_i + (1 - m) \cdot r_i,
\end{equation}
where \( \Pi_i \) denotes the cumulative payoff of individual \( i \) obtained through game interactions, \( r_i \) is the individual's current reputation, and \( m \in (0,1) \) is a reputation sensitivity factor that modulates the relative weight of material versus reputation components. A smaller value of \( m \) indicates a higher sensitivity to reputation, while a larger value emphasizes the increased weight of payoff.

Strategy evolution employs an asynchronous update mechanism. After each round of interactions, a randomly selected individual \( i \) chooses a neighbor \( j \) and evaluates the prospect of adopting \( j \)’s strategy. Here we use the Fermi update rule, which defines the probability of strategy adoption as:
\begin{equation}
P(S_i \leftarrow S_j) = \frac{1}{1 + \exp\left(\frac{f_i - f_j}{\kappa}\right)},
\end{equation}
where \( \kappa > 0 \) is the selection intensity or noise parameter that modulates the stochasticity of decision-making. When \( \kappa \) is large, the adoption probability becomes less sensitive to differences in fitness, implying greater randomness in behavioral transitions. Conversely, as \( \kappa \) approaches zero, individuals are more likely to deterministically imitate neighbors with higher fitness, thereby reinforcing selective pressure and accelerating the convergence of dominant strategies.

The details of our model are summarized in Fig.~\ref{fig1}. Each node in the network represents an agent, and the agent can choose between $C$ or $D$ in each round of interaction. Each agent has a reputation value that is dynamically updated over time, and by introducing an adaptive reputation threshold $\theta$, the agents are divided into two categories: high-reputation individuals (marked red) and low-reputation individuals (marked blue). The reputation threshold $\theta$ is adjusted as the average reputation of the group changes, thereby achieving dynamic evolution of the individual reputation state. The reputation attributes between agents have a decisive influence on the type of game they participate in: when two high-reputation individuals meet, a high-value game (referred to as Game~1) is played. When two low-reputation individuals interact, a low-value game (referred to as Game~2) is played, where the payoff is a fixed value $b$, which has nothing to do with reputation. If one of the two interacting parties is a high-reputation individual and the other is a low-reputation individual, the game type between the two parties is transferred to a high-value game with probability $p$, and to a low-value game with probability $(1-p)$, thus introducing a mechanism for the interference of reputation heterogeneity on the game structure. After each round of the game, the reputation value of the agent is updated according to the strategy choices of itself and its neighbors, promoting coordinated changes of the strategy evolution and reputation structure.

\section{Results}
\label{sec: results}
Next, we analyze the evolution of cooperation and the reputation threshold under different game parameters. To account for the potential impact of network structure on system dynamics, we both use square lattice and Watts–Strogatz small-world networks. In the latter case, an average degree $k = 10$ and the rewiring probability $p = 0.5$ are used. For comparison, all networks contain $N = 2500$ nodes. Initially, each agent is randomly assigned to be a cooperator or a defector with equal probability. All simulations are implemented using Python 3.10.

\subsection{Heatmap of Cooperation Density}
\begin{figure*}[htbp]
    \centering
    \begin{subfigure}[t]{0.48\linewidth}
        \centering
        \includegraphics[width=\linewidth]{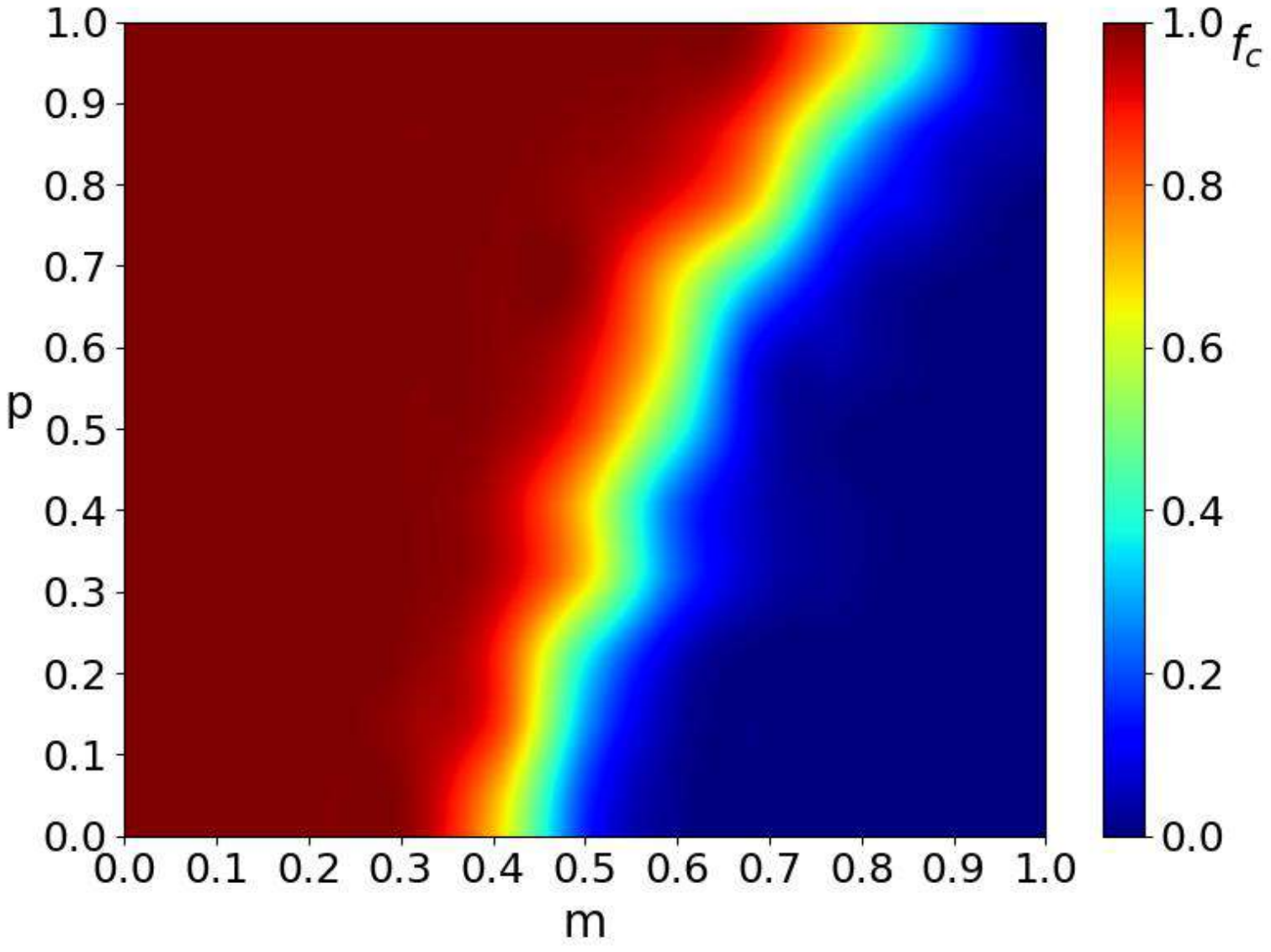}
        \caption{SL}
    \end{subfigure}
    \hfill
    \begin{subfigure}[t]{0.48\linewidth}
        \centering
        \includegraphics[width=\linewidth]{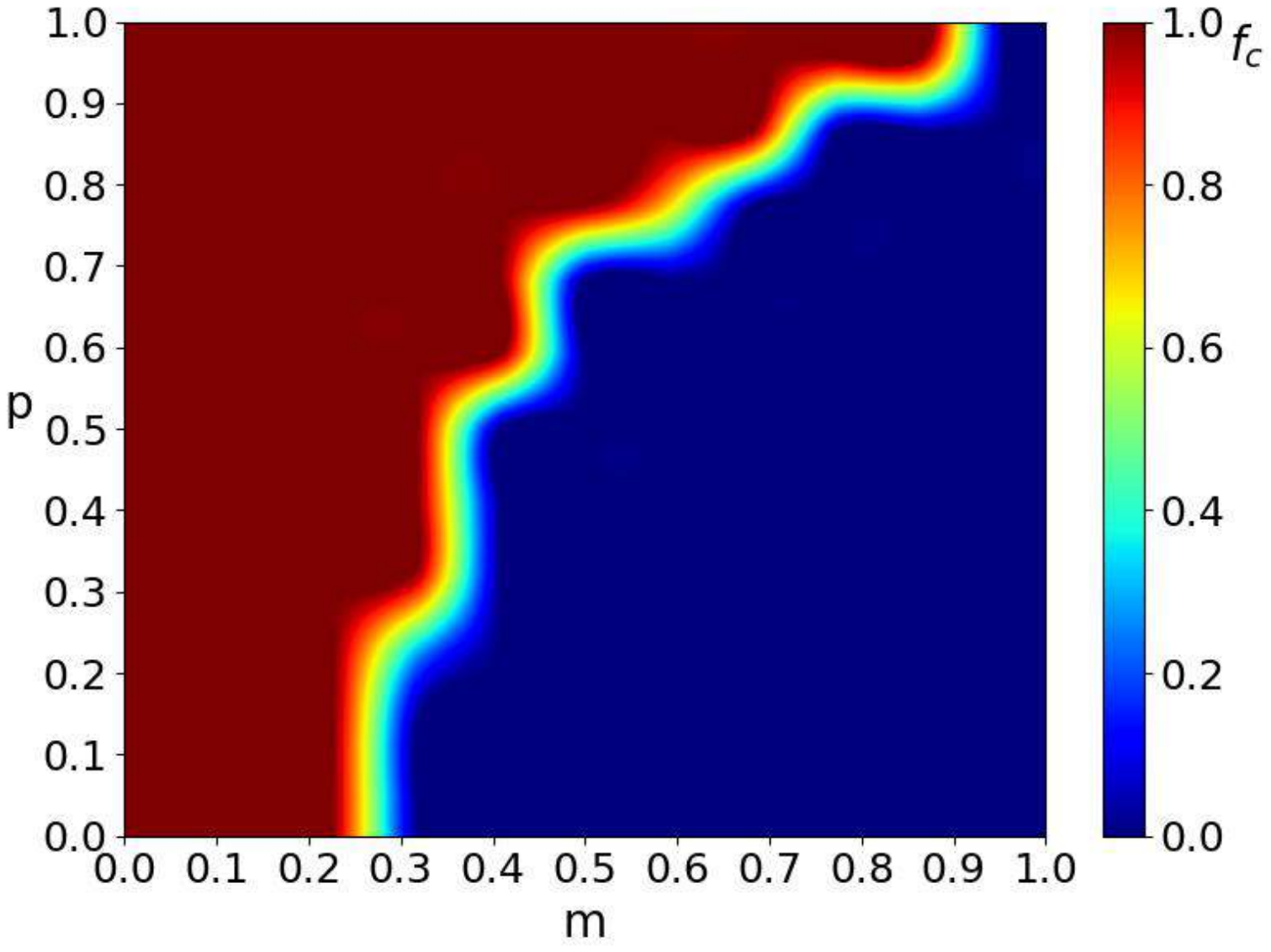}
        \caption{WS}
    \end{subfigure}
    \caption{\footnotesize \textbf{Cooperation level on the $p-m$ parameter plane at different networks.} Color-coded cooperation density $f_c$ is shown in dependence of the game asymmetry parameter $p$ and the reputation sensitivity factor $m$ for square lattice (SL) and for small-world (WS) network. The value of parameter $b_l=1.1$ is fixed for both cases. The actual value of $f_c$ is explained in the legend to the right of each panel.}
    \label{fig2}
\end{figure*}

By systematically adjusting the relative weight factor $m$ and the game transition probability $p$ in the extended fitness function, we further examined the evolutionary characteristics of the cooperation level in the system. To comprehensively evaluate the universality of the system behavior and the impact of the network structure on the evolution process, we conducted simulation experiments in the two typical network structures mentioned as SL and WS. Fig.~\ref{fig2} summarizes the simulation results, in which the cooperation cost $c = 1$ and the temptation benefit $b_l = 1.1$ of the low-value game are uniformly set in all cases. From the results, it can be observed that both in the regular lattice and the small-world network, with the decrease of the game transition probability $p$ and the increase of the weight factor $m$, the overall cooperation density of the system shows a significant decline. In addition, within the parameter range examined, we can clearly divide the parameter areas corresponding to pure cooperation and pure defection. It is worth noting that the phase transition process from cooperation to defection shows an obvious right-skewed trend under different network structures, indicating that there are consistent phase transition characteristics in game dynamics and that the network structure has a non-trivial modulating effect on the transition process.

In addition, we further observed that the network topology plays a crucial regulatory role in the process of cooperation evolution. Specifically, in the SL and WS, when the relative weight factor \( m \) reaches about \( 0.4 \) and \( 0.25 \) respectively, the density of cooperators drops sharply, forming a clearly visible transition line on the \( p \)-\( m \) parameter plane. The phenomenon indicates that the system has undergone a transition from a cooperation-dominated state to a defection-dominated state, that is, there is an obvious critical point for cooperation evolution. In particular, in the regular network, the system has a higher tolerance for changes in \( p \) and \( m \), and the transition region is wider and smoother than that of WS. The structural feature provides a greater possibility for the long-term coexistence of cooperation and defection strategies, delays the trend of pure defection strategies that dominate the whole population, and effectively suppresses the occurrence of a complete collapse of cooperation. In contrast, due to the rapid spread of information caused by higher connectivity and short average path length, the system is more sensitive to disturbances, and the critical behavior of cooperation turning into defection is sharper and steeper in WS networks. In other words, the network structure not only affects the overall level of cooperation in the system but also determines the robustness and fragility of cooperation stability. The structure of the regular grid helps to maintain the existence of the early cooperation groups through localized interaction restrictions, providing more living space for cooperation. While the WS effect promotes information diffusion, it also exacerbates instability in the process of strategy evolution.

\subsection{The Formation of Cooperation-Dominated State Under Different $b_l$}
\begin{figure}[htbp]
    \centering
    \includegraphics[width=0.5\linewidth]{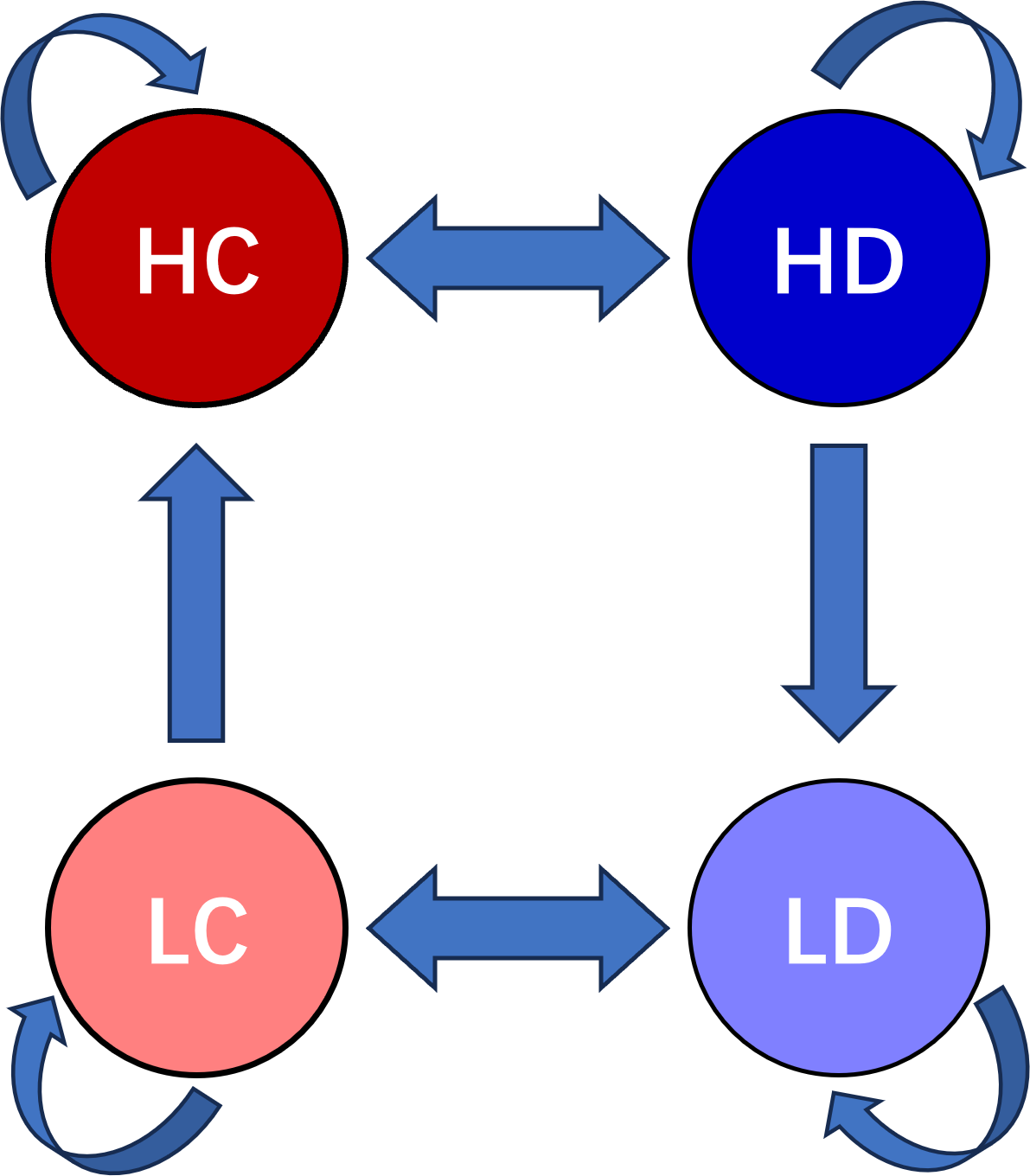}
    \caption{\textbf{Schematic diagram of possible strategy-reputation states.} Red represents high reputation cooperators (HC), dark blue represents high reputation defectors (HD), pink represents low reputation cooperators (LC), and light blue represents low reputation defectors (LD), showing the possible transition of four possible states of individuals in the framework of strategy and reputation.}
    \label{fig3}
\end{figure}

\begin{figure*}[htbp]  
    \centering
    \newcolumntype{C}{@{}>{\centering\arraybackslash}p{0.23\linewidth}@{}}
    
    \begin{tabular}{@{}lCCCC@{}}
        & \textbf{step=0} & \textbf{step=500} & \textbf{step=2000} & \textbf{step=3000} \\
        \vspace{-3ex} 
    \end{tabular}
    
    \begin{tabular}{@{}lCCCC@{}}
        \raisebox{3\height}{\rotatebox[origin=c]{90}{\textbf{$b_l=1.1$}}} & 
        \includegraphics[width=\linewidth]{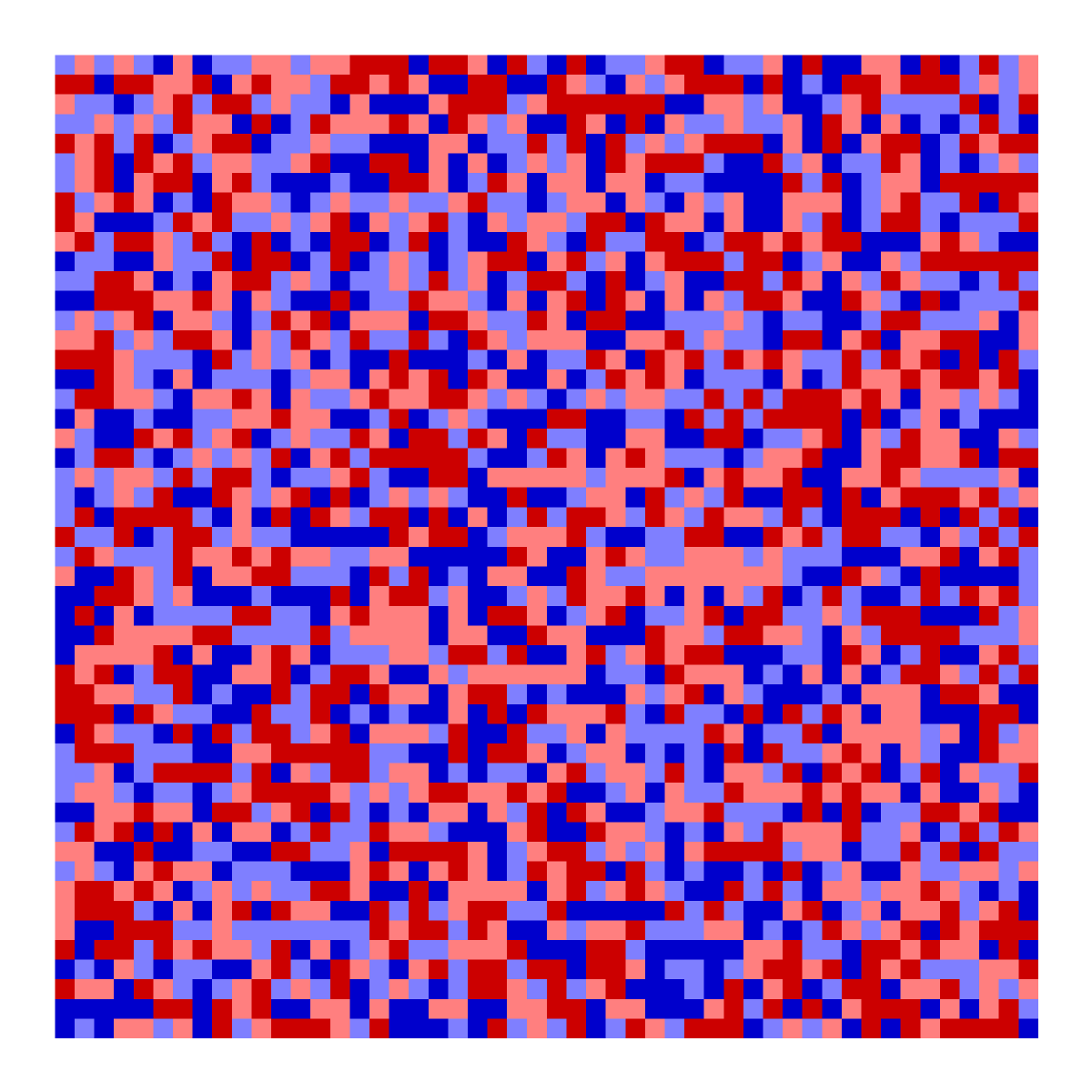} &
        \includegraphics[width=\linewidth]{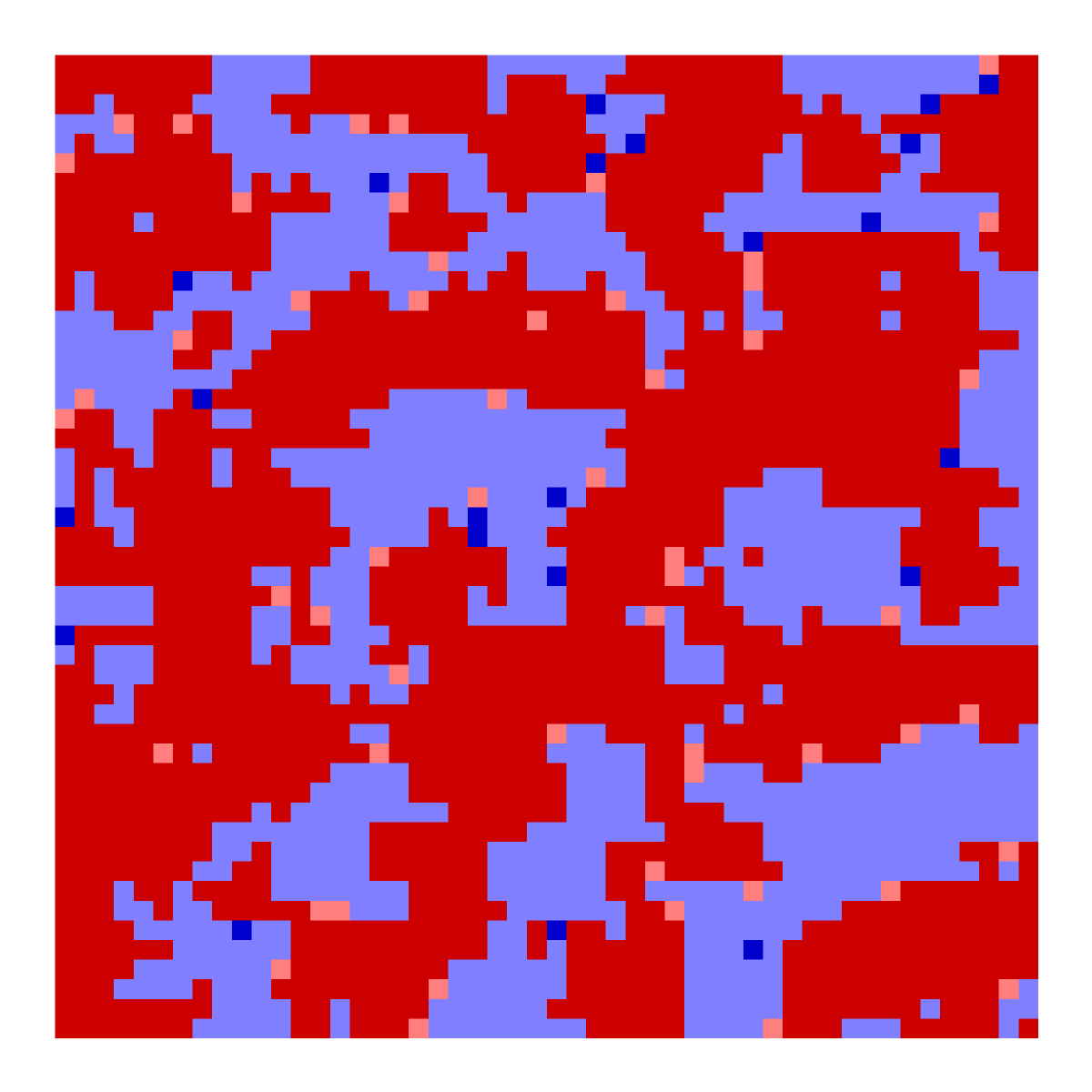} &
        \includegraphics[width=\linewidth]{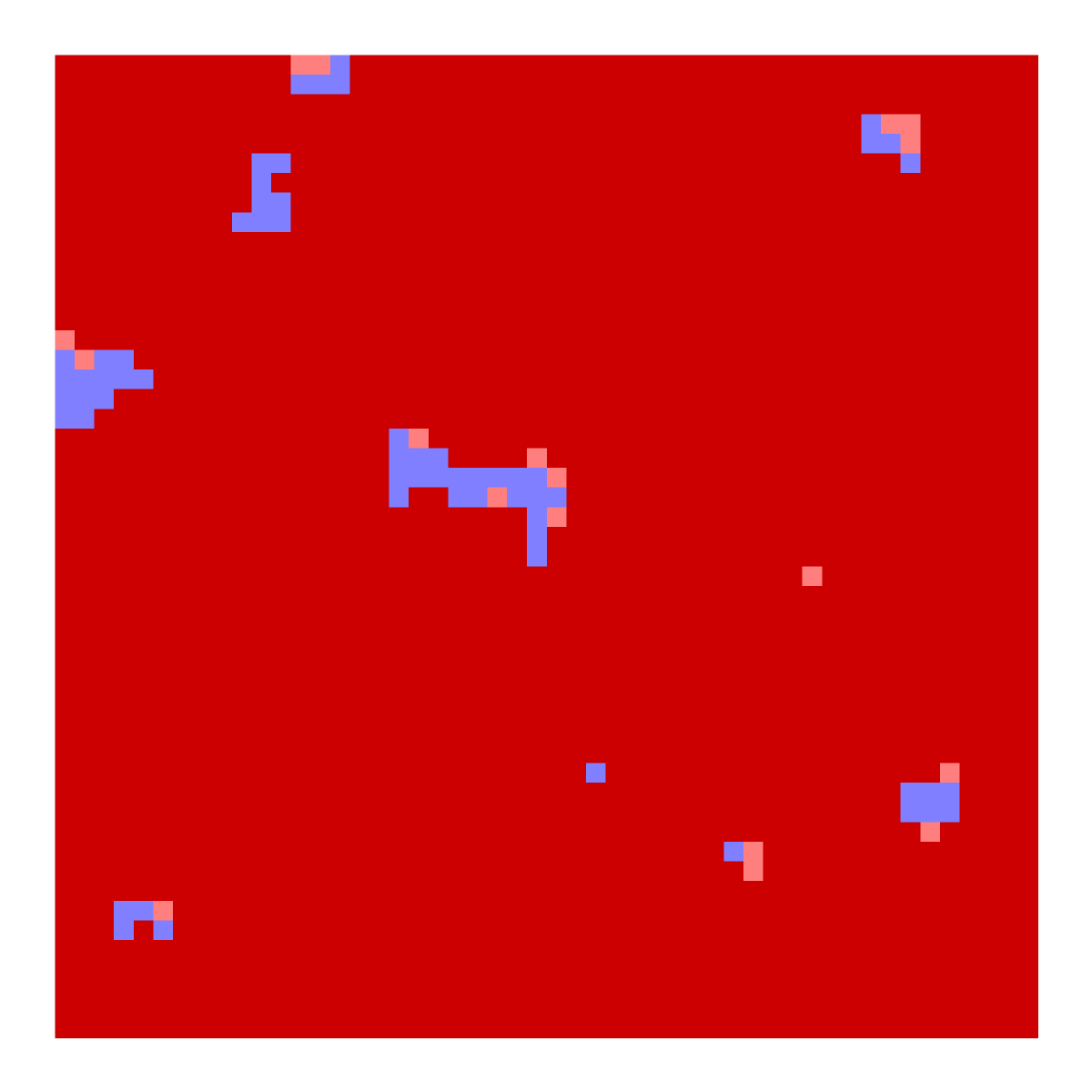} &
        \includegraphics[width=\linewidth]{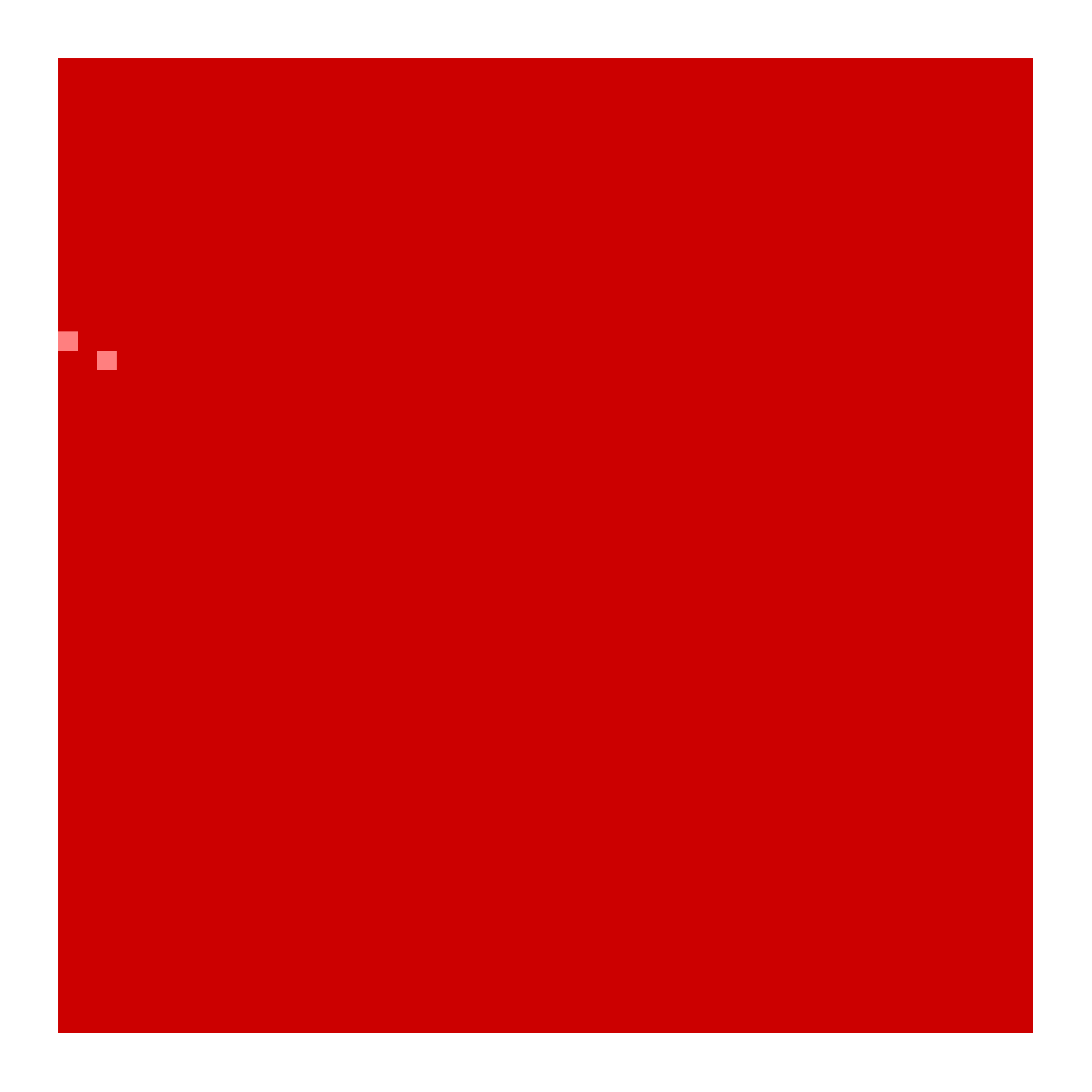} \\
    \end{tabular}
    
    \vspace{0.cm} 
    
    \begin{tabular}{@{}lCCCC@{}}
        \raisebox{3.5\height}{\rotatebox[origin=c]{90}{\textbf{$b_l=1.5$}}} &
        \includegraphics[width=\linewidth]{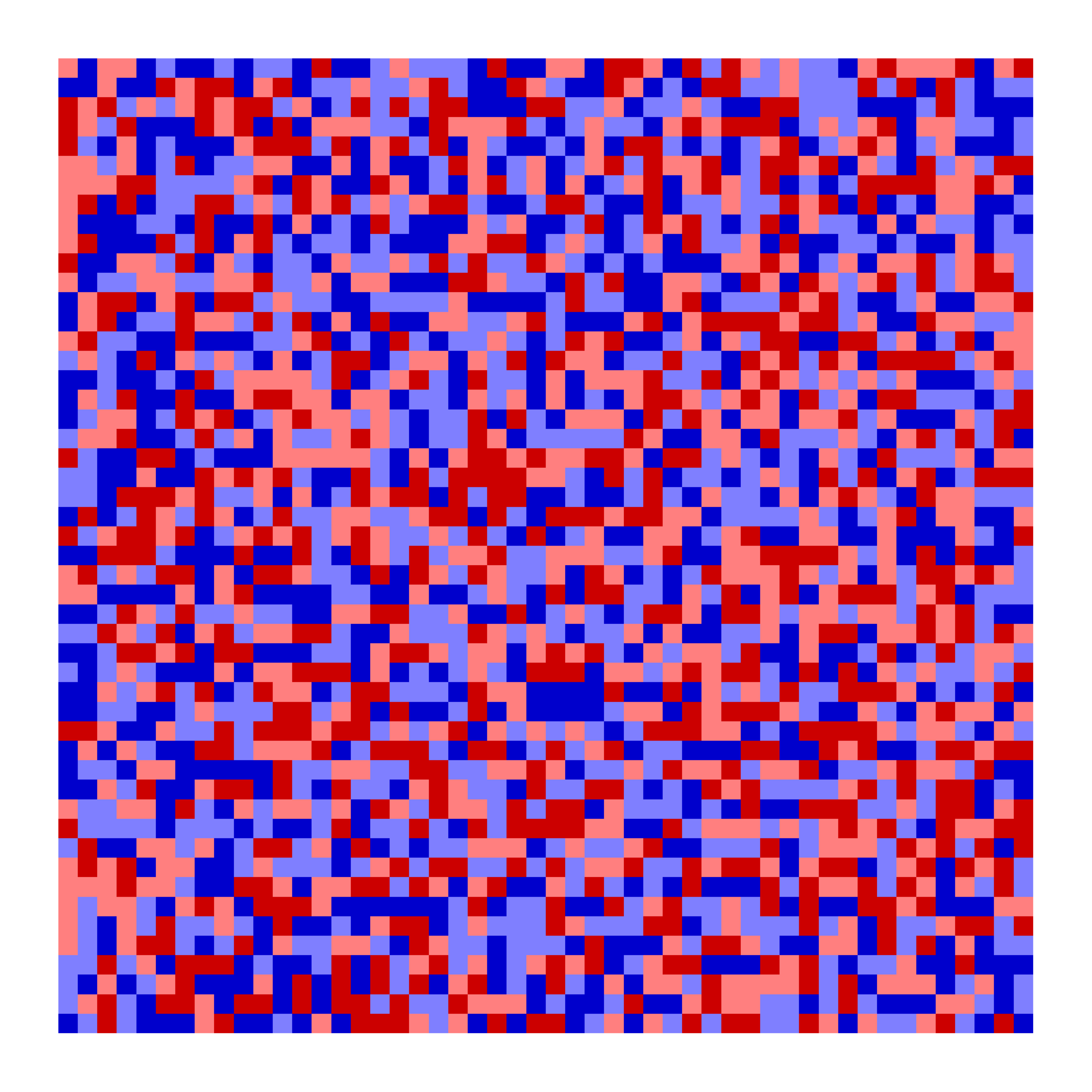} &
        \includegraphics[width=\linewidth]{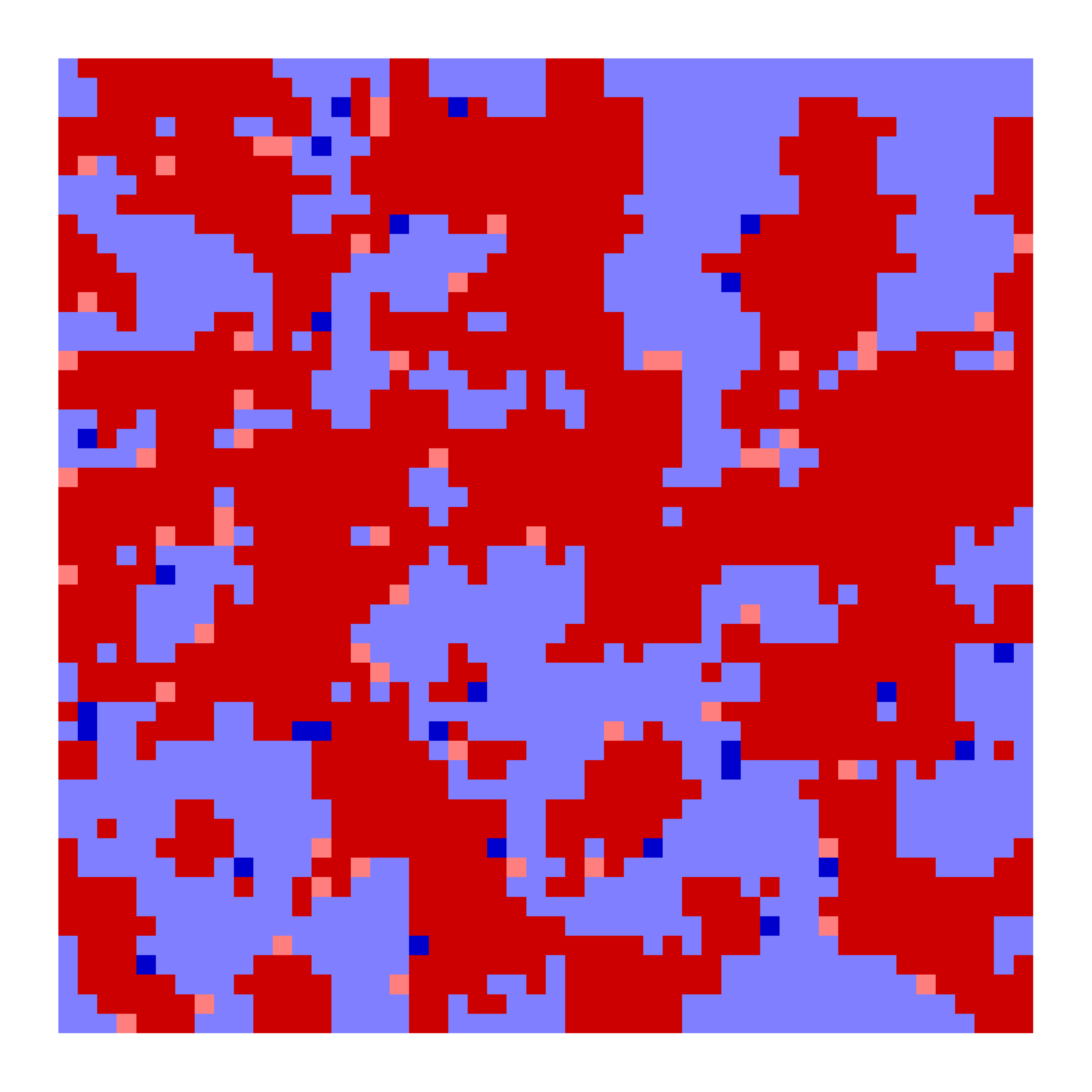} &
        \includegraphics[width=\linewidth]{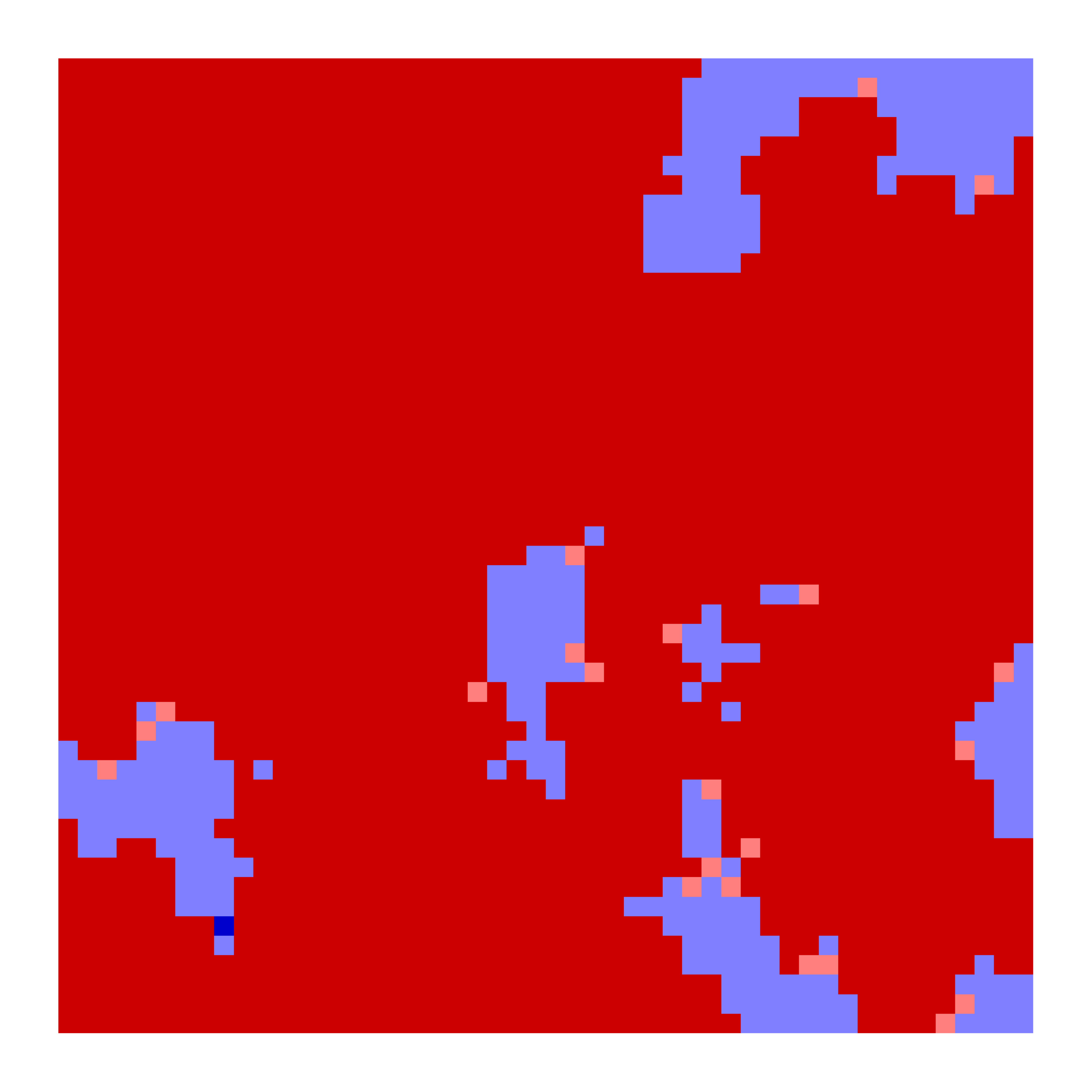} &
        \includegraphics[width=\linewidth]{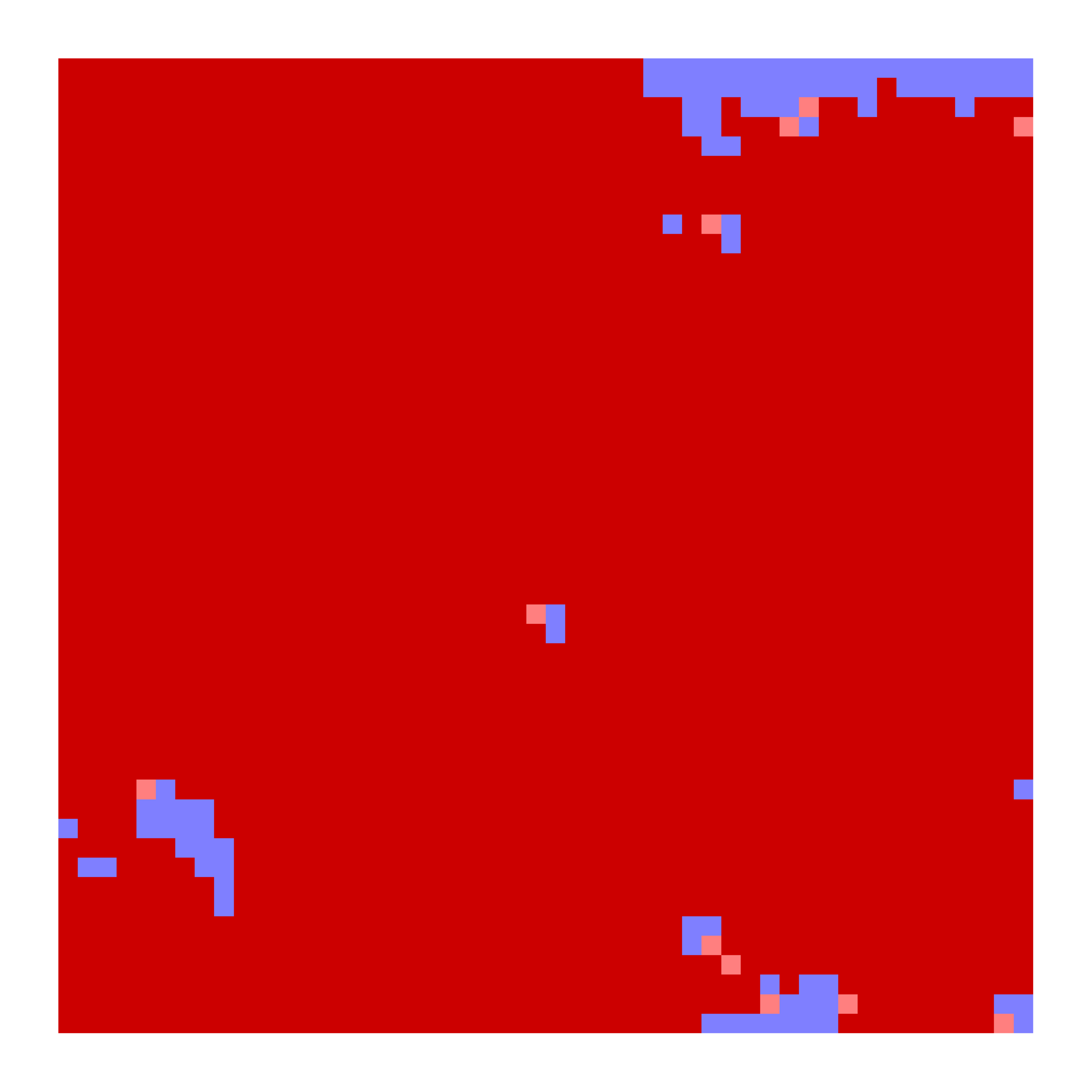} \\
    \end{tabular}
    
    \vspace{0.05cm}
    
    \begin{tabular}{@{}lCCCC@{}}
        \raisebox{4.5\height}{\rotatebox[origin=c]{90}{\textbf{$b_l=2$}}} &
        \includegraphics[width=\linewidth]{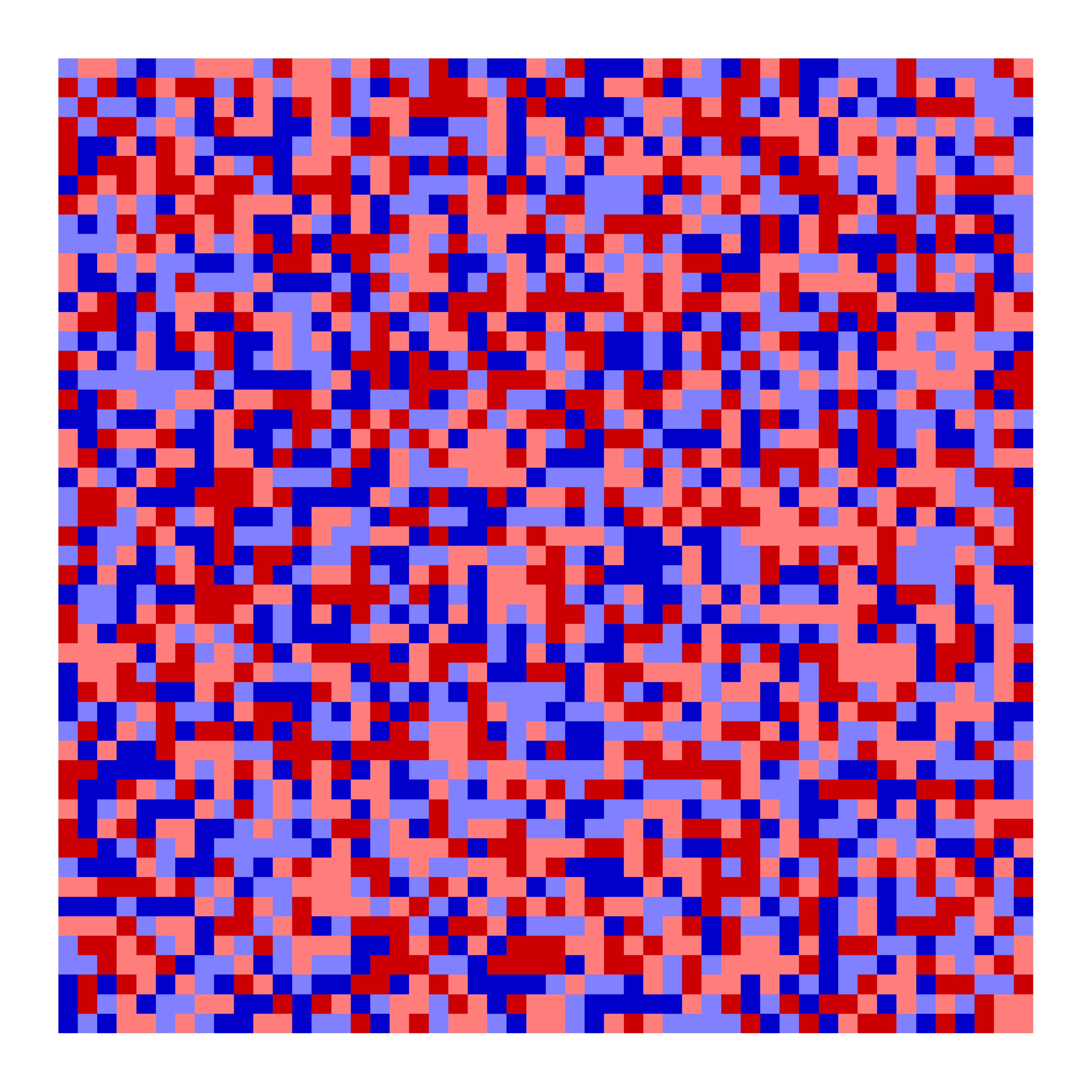} &
        \includegraphics[width=\linewidth]{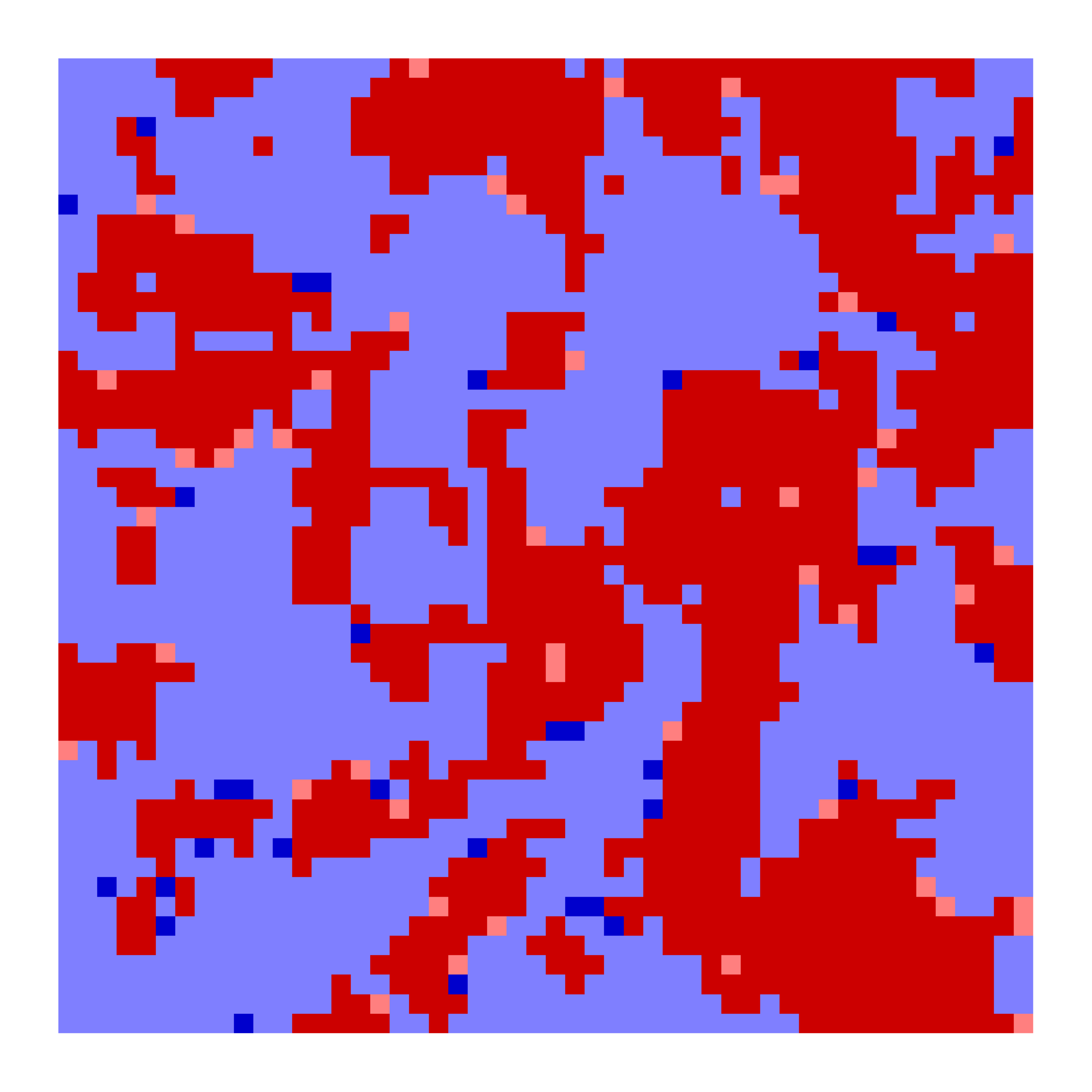} &
        \includegraphics[width=\linewidth]{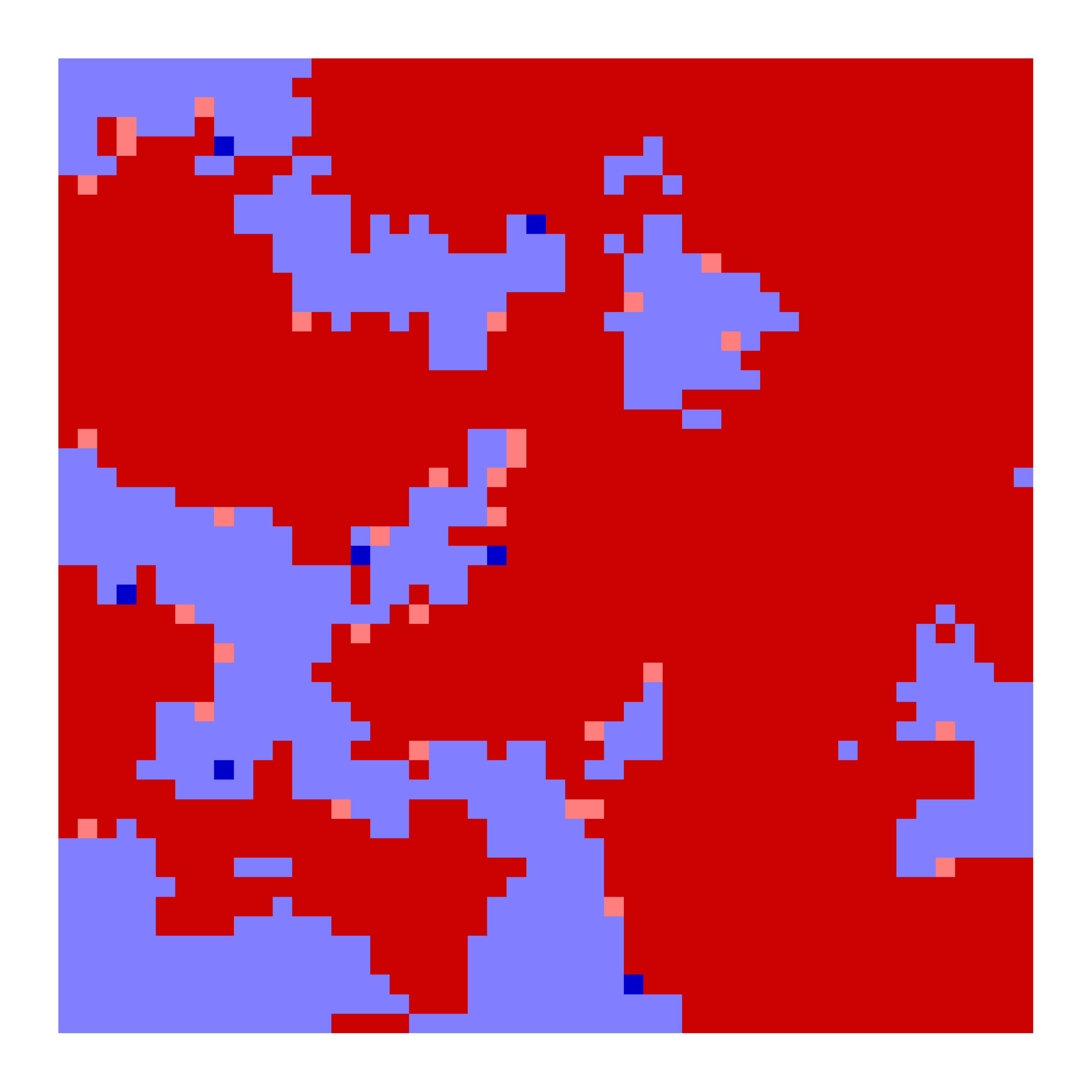} &
        \includegraphics[width=\linewidth]{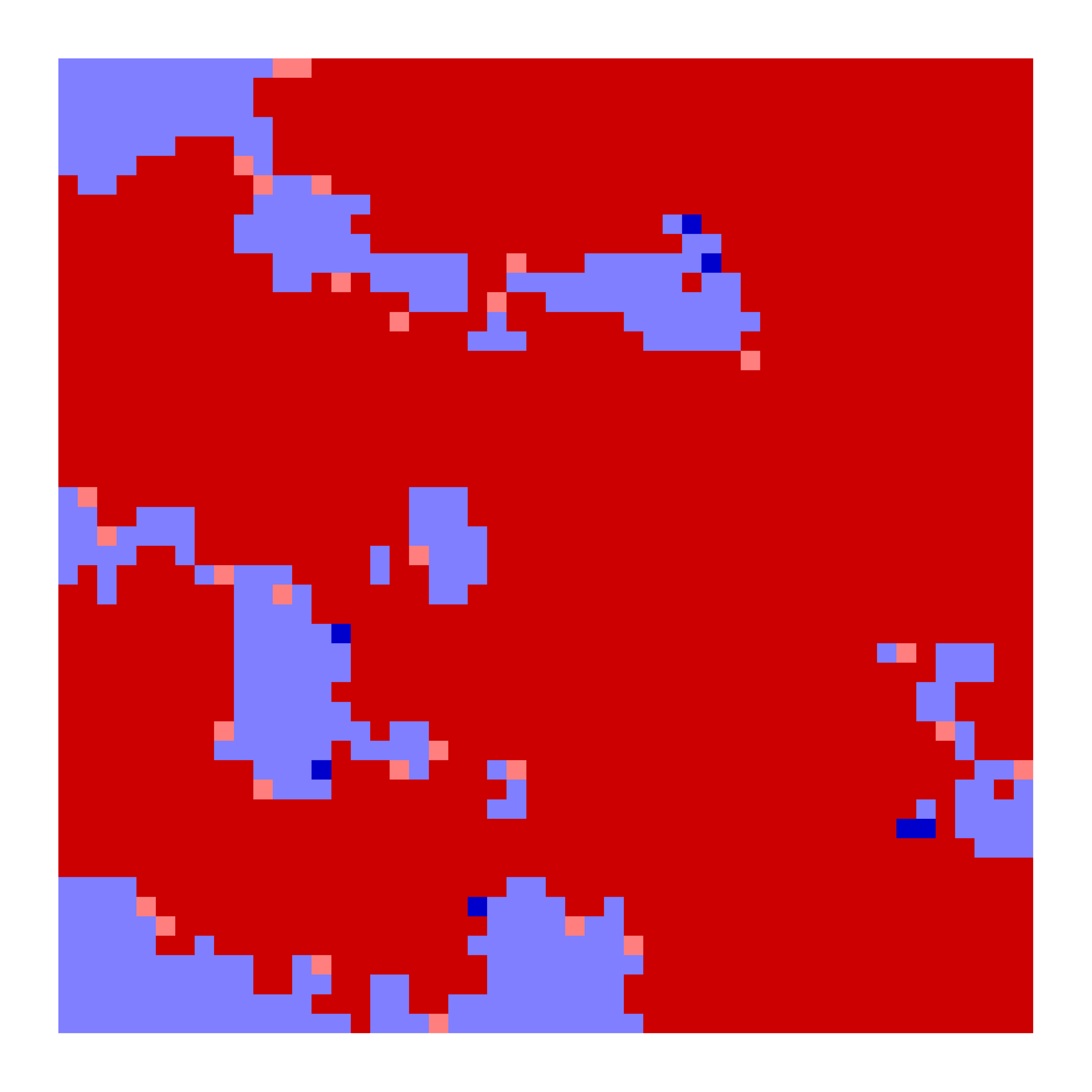} \\
    \end{tabular}
    
   \caption{\textbf{Spatial evolution of individual states on SL under different temptation intensities.} From top to bottom, the rows correspond to \( b_l = 1.1 \), \( 1.5 \), and \( 2.0 \). From left to right, the columns show the spatial distribution of states at four representative stages of the evolutionary process (obtained at MCS = 0, 500, 2000, and 3000 steps). All simulations are performed under fixed parameters \( m = 0.6 \) and \( p = 0.9 \). The color-code of players, which informs about the individual strategy and reputation, is identical to those we used in Fig.~\ref{fig3}. }
    \label{fig4}
\end{figure*}

After introducing the reputation mechanism to regulate the heterogeneous game structure, it is worth noting whether the reputation-based classification mechanism can effectively promote the continuous evolution of cooperation, especially when multiple game types coexist. When low-value games exert strong strategic inducements, individuals may opt for defection strategies driven by short-term profit incentives. In such scenarios, the ability of the reputation mechanism to maintain a steady state dominated by cooperation through dynamic constraints becomes a critical criterion for evaluating its effectiveness and robustness. In this context, we analyze the long-term evolutionary dynamics of the model under varying levels of temptation in low-value games, denoted by \( b_l \), to investigate how reputation facilitates cooperation within heterogeneous game environments. 

To facilitate the analysis of individual state transitions during the strategy evolution, we first illustrate the four possible states of individuals in Fig.~\ref{fig3}, categorized by their strategies (C/D) and reputation levels (high/low). Specifically, red nodes represent high-reputation cooperators (HC), dark blue nodes represent high-reputation defectors (HD), pink nodes correspond to low-reputation cooperators (LC), and light blue nodes represent low-reputation defectors (LD). Using this color-coding, Fig.~\ref{fig4} presents the evolution of the spatial distribution of individual status under fixed parameters \( p = 0.9 \) and \( m = 0.5 \), with varying values of \( b_l \). We consider three representative cases: \( b_l = 1.1 \), \( 1.5 \), and \( 2.0 \), which correspond to the upper, middle, and lower rows of Fig.~\ref{fig4}, respectively. Each snapshot presents a typical configuration of the system at a given time step. 

Simulation results demonstrate that even under high levels of temptation in low-value games, cooperators can still proliferate widely and become dominant, while defection is substantially suppressed. Notably, cooperation is predominantly concentrated among high-reputation individuals, whereas defection is mainly observed among low-reputation individuals. These patterns reveal a coupling feedback mechanism between strategy and reputation: reputation functions not only as a determinant of game type selection but also imposes structural constraints on the diffusion of strategies. A stable correlation between high-reputation cooperation and low-reputation defection emerges, driving the system toward a cooperation-dominated equilibrium.

\subsection{Effect of Reputation Sensitivity on the Steady-State Reputation Threshold}
\begin{figure*}[htbp]
    \centering
    \begin{subfigure}[t]{0.48\linewidth}
        \centering
        \includegraphics[width=\linewidth]{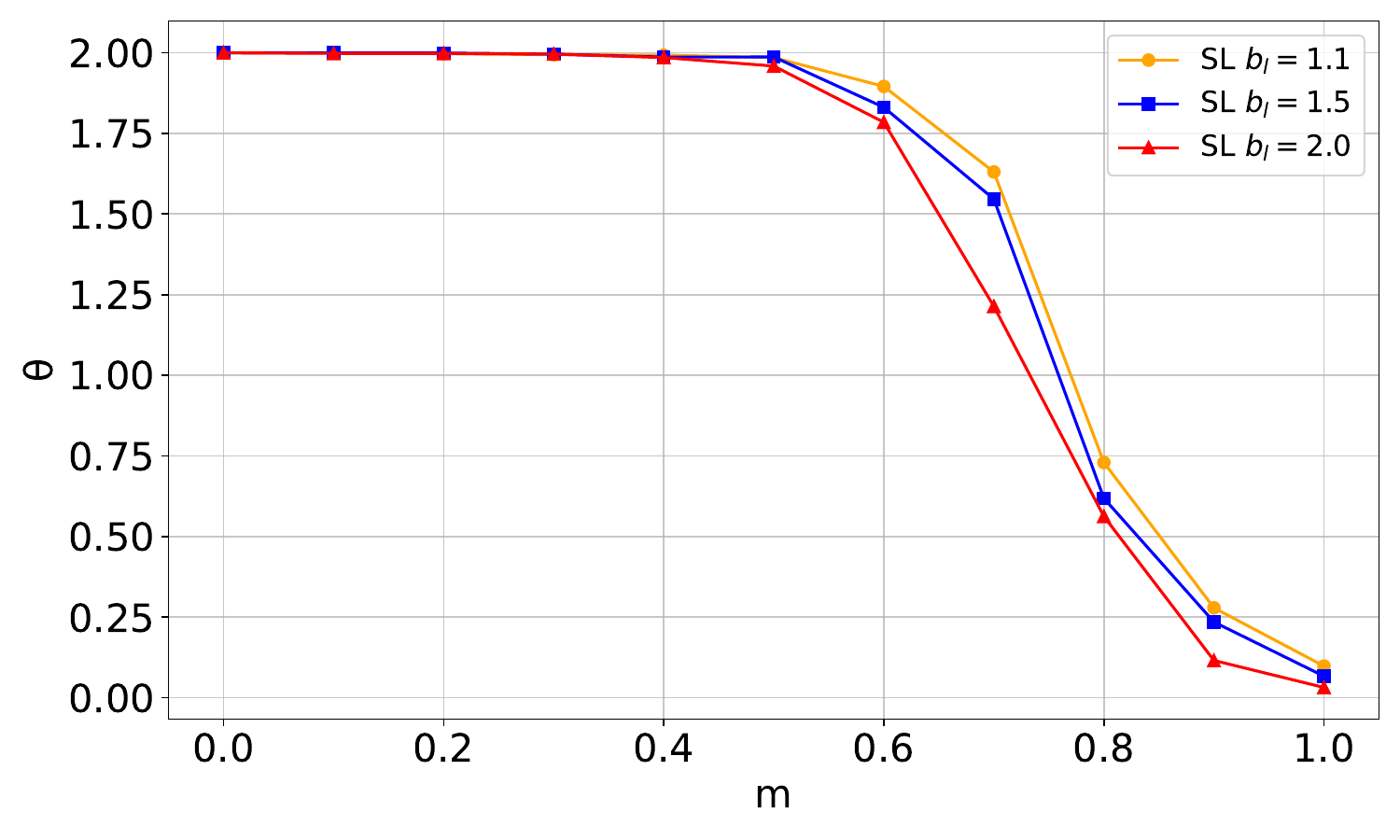}
        \caption{SL}
        \label{fig:2a}
    \end{subfigure}
    \hfill
    \begin{subfigure}[t]{0.48\linewidth}
        \centering
        \includegraphics[width=\linewidth]{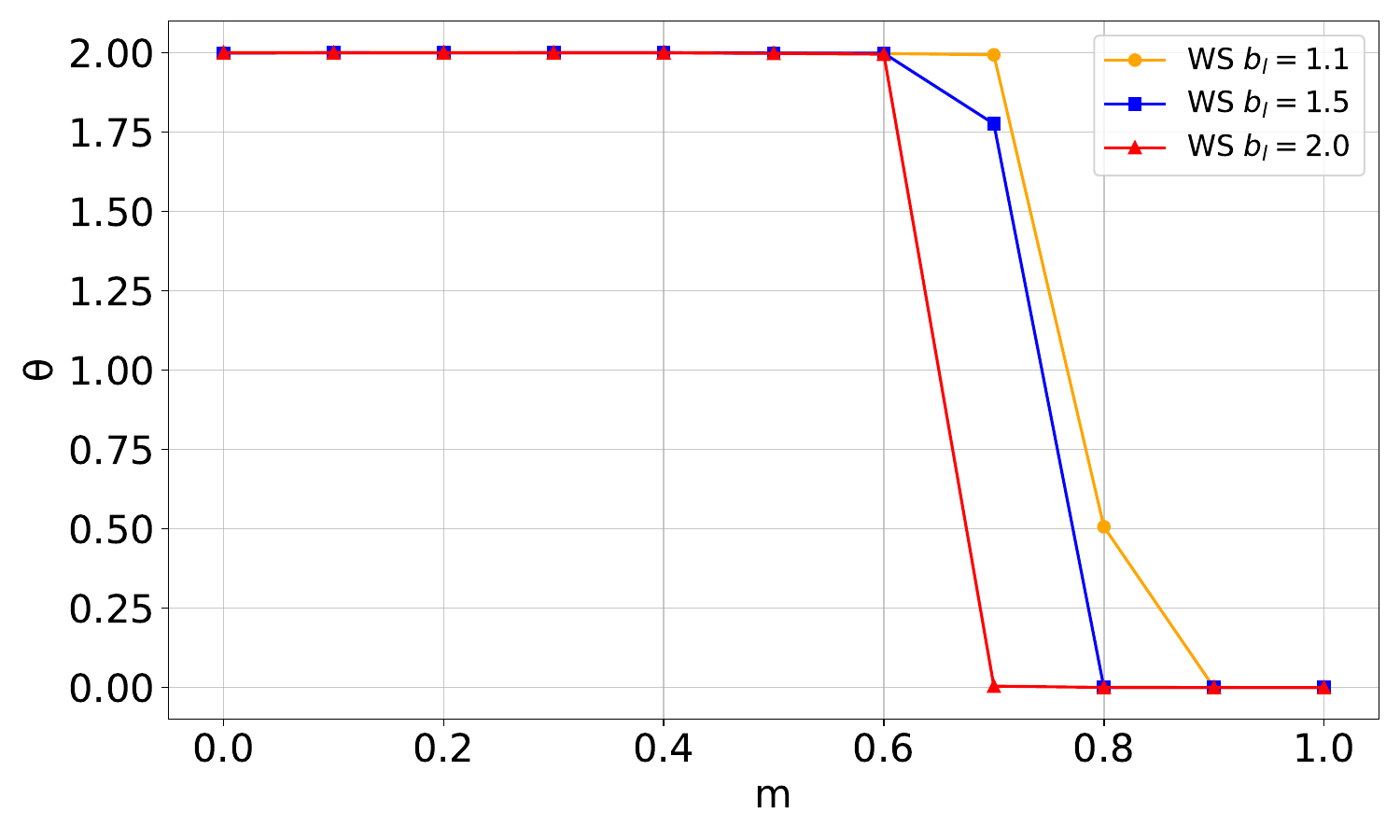}
        \caption{WS}
        \label{fig:2b}
    \end{subfigure}
    \caption{\footnotesize \textbf{The steady-state reputation threshold of the system under different combinations of \( b_l \) and \( m \) parameters.} The three colors in the figure correspond to different temptation values, as shown in the legends. Panel~(a) illustrates the variation of the reputation threshold concerning \( m \) on SL, whereas panel~(b) presents the corresponding results on WS network. The game transition probability is fixed \( p = 0.9 \) for both cases. }
    \label{fig5}
\end{figure*}

In a reputation-driven heterogeneous game system, the effectiveness of the reputation mechanism depends not only on how it guides strategy selection but also on the degree of influence of reputation itself on individual fitness. In our model, the influence is controlled by the parameter \( m \) which reflects the sensitivity of individuals to the reputation in the strategy evolution. Therefore,  it is of great significance to reveal how the reputation mechanism maintains cooperation under different temptation values to understand the relationship between the parameter \(m\) and the steady-state reputation threshold of the system. In particular, when the temptation of low-value games is strong, if individuals are insensitive to reputation influence (i.e., \( m \) is large), whether the system can still maintain an effective reputation differentiation and behavior regulation mechanism becomes the key to verifying the robustness of the model.

To elaborate on this issue, we systematically examined the relationship between the reputation weight factor \( m \) and the steady-state reputation threshold of the system under three typical low-value game temptation intensities (\( b_l = 1.1, 1.5, \) and \( 2.0 \)), and the results are shown in Fig.~\ref{fig5}. It can be clearly seen that the reputation threshold decreases significantly with increasing \( m \) at all \( b_l \) values. The trend shows that when the weight of reputation in individual fitness decreases, the individual's sensitivity to its reputation level decreases, making it easier to adopt defection, which leads to a continuous decline in reputation.

By further comparing the evolutionary characteristics under different network structures, we find that the network structure has an important influence on the robustness of the reputation mechanism. In the lattice, the reputation threshold shows a relatively gentle downward trend with an increase of \(m\), showing a strong structural buffering effect. In the WS network, however, the threshold decreases significantly faster, especially when \(m\) is between 0.7 and 0.8, the reputation thresholds corresponding to the three temptation intensities drop sharply, indicating that the reputation mechanism is ineffective under the condition. In other words, the effectiveness of the reputation mechanism in regulating the evolution of individual strategies is not only affected by the internal behavioral parameters of the individual but is also highly sensitive to the complexity of the network structure. Especially in a dynamic game environment, if the influence of reputation on fitness is insufficient, it will be difficult to support the continued role of its strategy regulation mechanism, thereby weakening the occurrence of cooperation.

\subsection{Robustness of cooperation Evolution under Diverse Initial Reputation Distributions}

\begin{figure*}[htbp]
    \centering

    \begin{subfigure}[t]{0.48\linewidth}
        \centering
        \includegraphics[width=\linewidth]{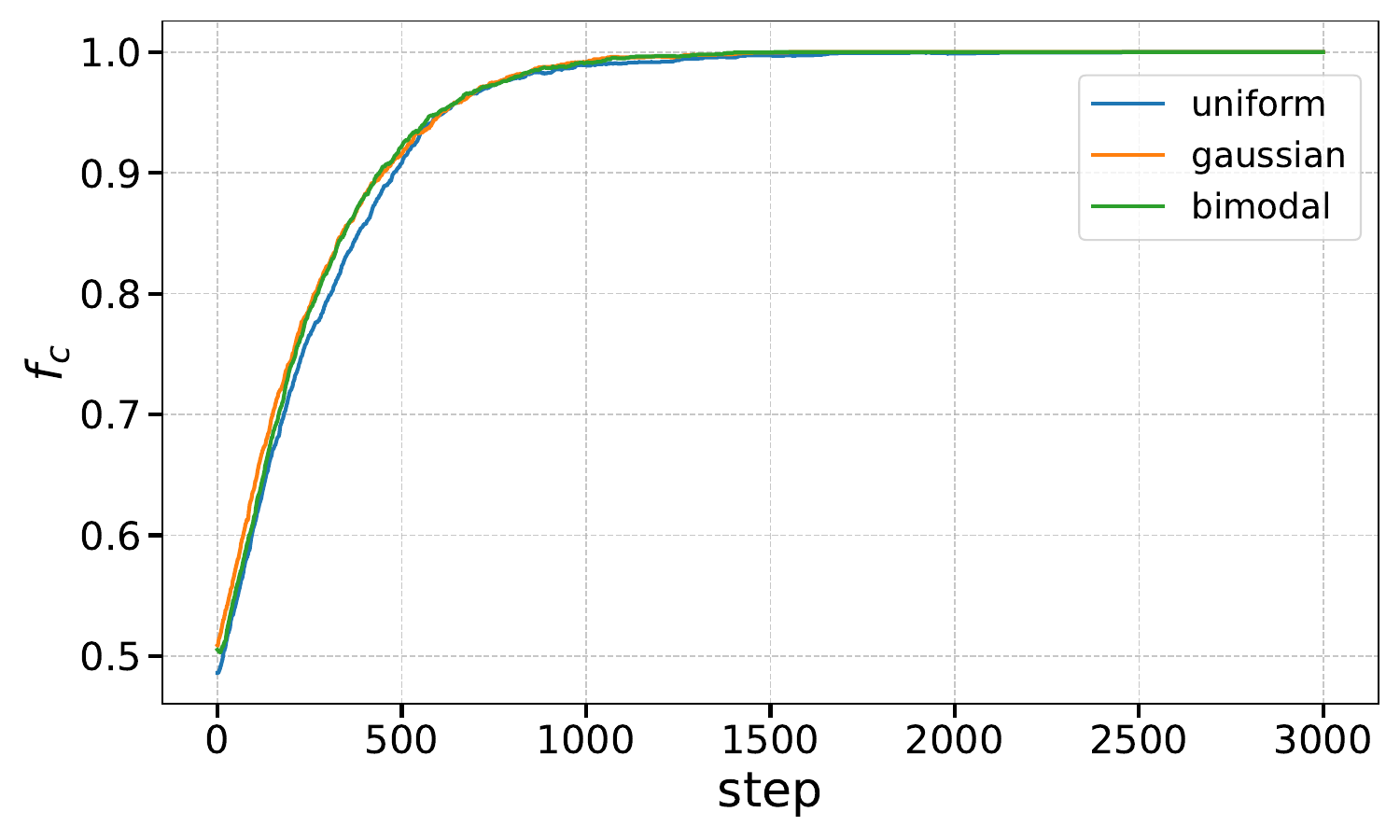}
    \end{subfigure}
    \hfill
    \begin{subfigure}[t]{0.48\linewidth}
        \centering
        \includegraphics[width=\linewidth]{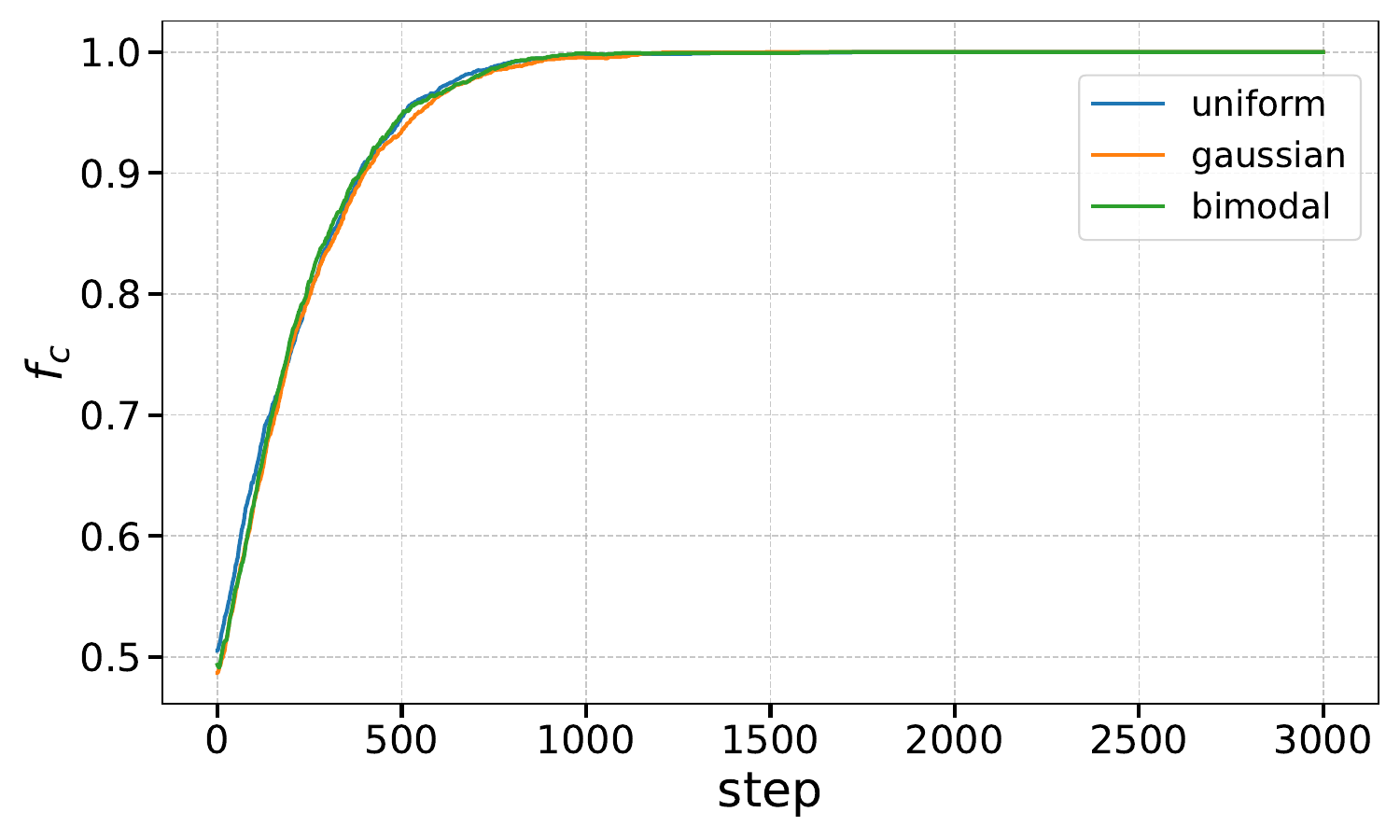}
    \end{subfigure}
    
    \vspace{0.2cm}

    \begin{subfigure}[t]{0.48\linewidth}
        \centering
        \includegraphics[width=\linewidth]{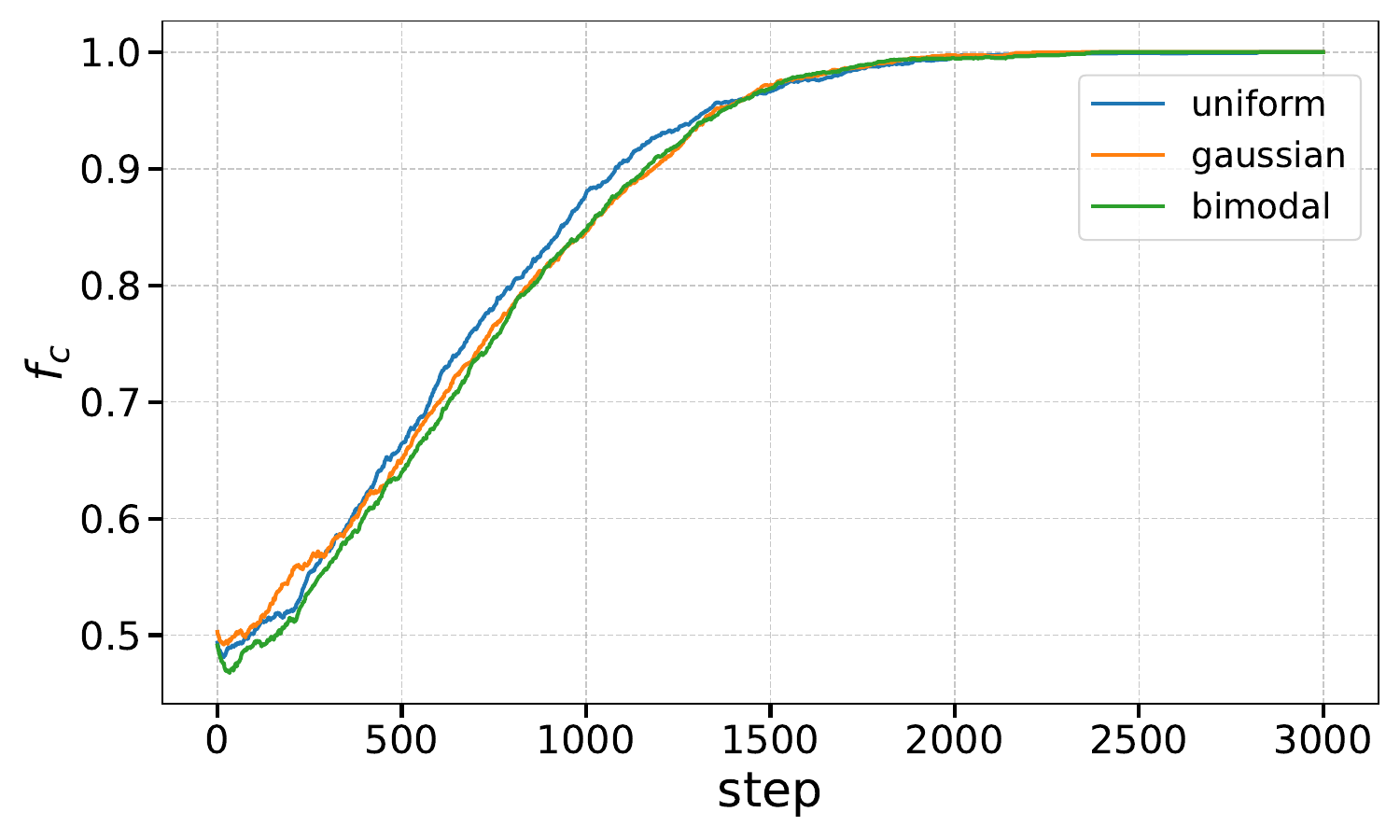}
    \end{subfigure}
    \hfill
    \begin{subfigure}[t]{0.48\linewidth}
        \centering
        \includegraphics[width=\linewidth]{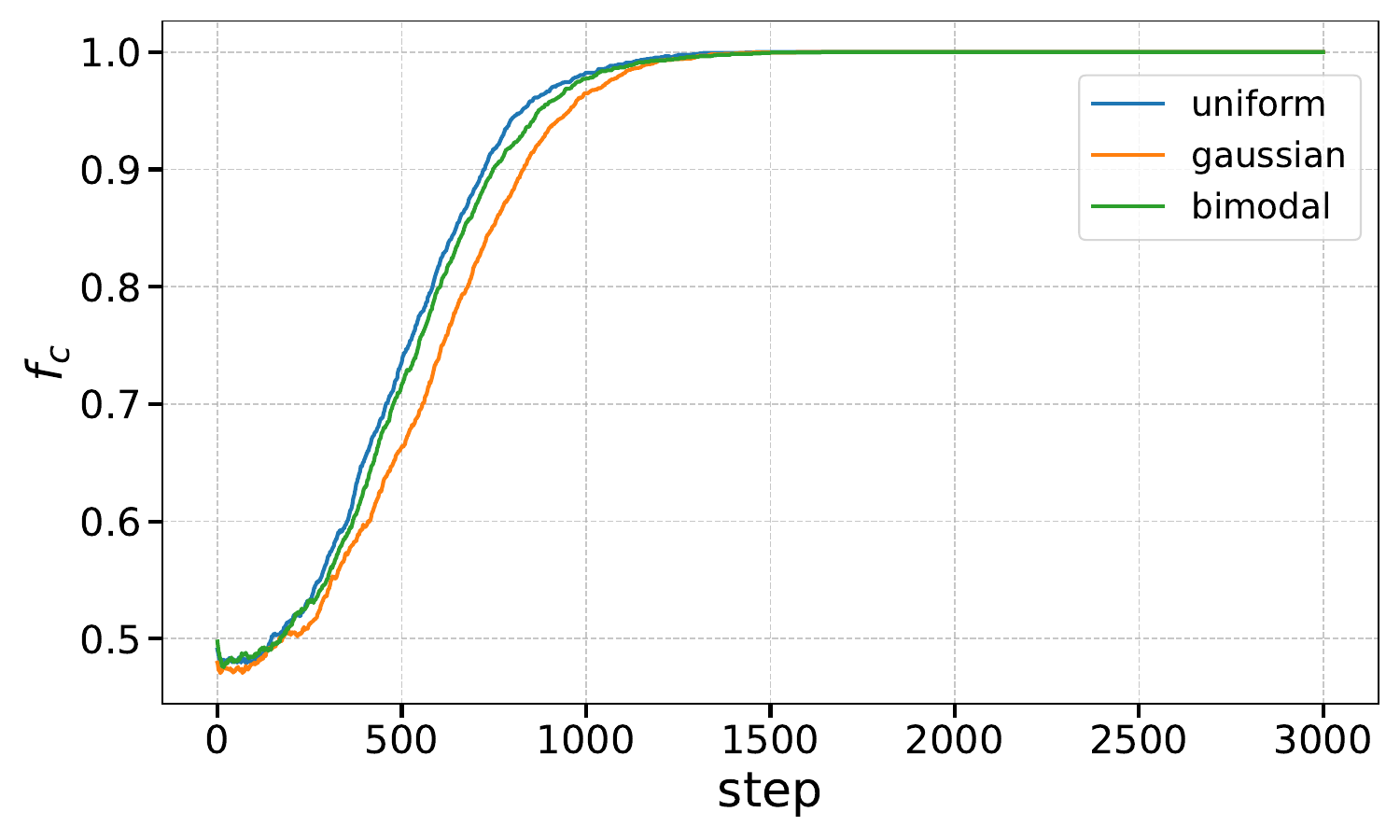}
    \end{subfigure}
    
    \vspace{0.2cm}

    \begin{subfigure}[t]{0.48\linewidth}
        \centering
        \includegraphics[width=\linewidth]{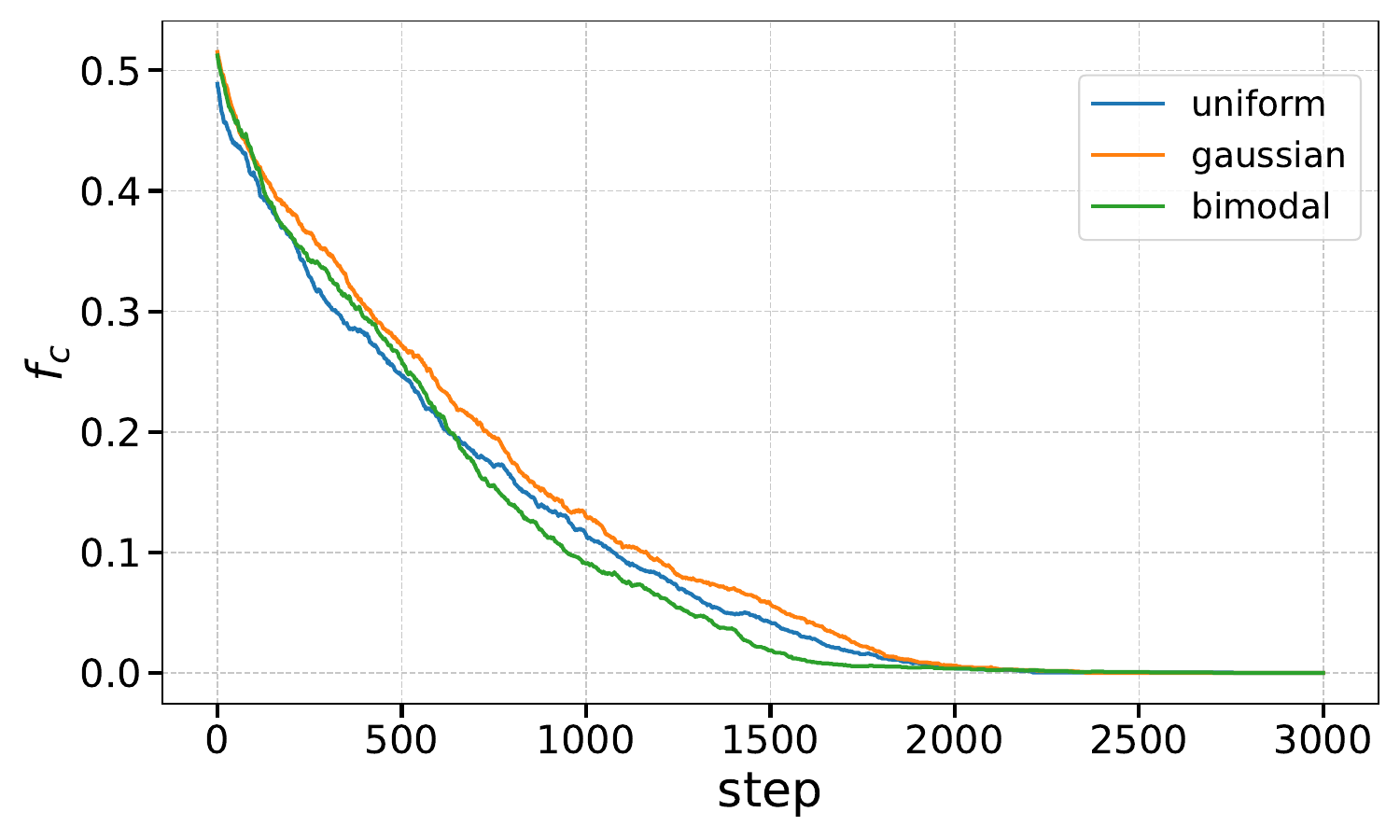}
    \end{subfigure}
    \hfill
    \begin{subfigure}[t]{0.48\linewidth}
        \centering
        \includegraphics[width=\linewidth]{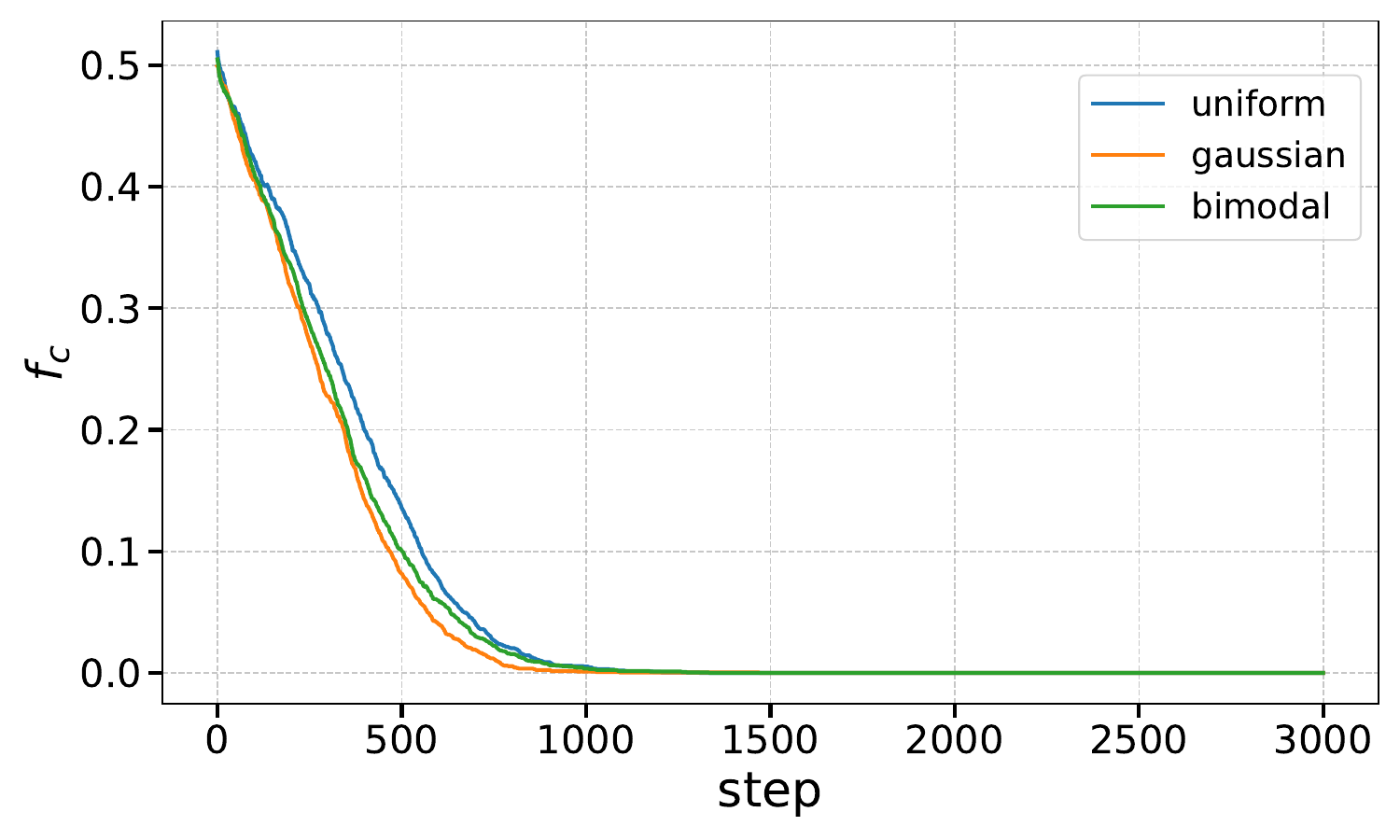}
    \end{subfigure}

    \caption{\textbf{The evolution of cooperation level under different initial reputation distributions.} Each row corresponds to a specific value of the reputation sensitivity parameter \(m\) (from top to bottom: \(m = 0\), \(0.5\), and \(1.0\)). The left column presents simulation results on SL, while the right column shows results on WS graph. Each panel compares three types of initial reputation distributions: uniform, Gaussian, and bimodal. All simulations are conducted with fixed parameters: \(p = 0.9\) and \(b_l = 1.5\).}

    \label{fig6}
\end{figure*}

In a reputation-driven evolutionary game system, while the endogenous update mechanism of reputation is central to the long-term dynamics, the initial conditions—particularly the distribution of initial reputations—may exert a non-negligible influence on the early-stage formation and propagation of cooperation structures. Real-world social systems often exhibit heterogeneous reputation landscapes from the outset. For instance, some communities are characterized by a few individuals with exceptionally high reputations, while others demonstrate more uniform or multimodal distributions. The heterogeneity necessitates a systematic investigation of how different initial reputation patterns condition the trajectory and robustness of cooperation.

To clarify this, we construct three representative scenarios of initial reputation distributions: a uniform distribution, a Gaussian (normal) distribution, and a bimodal distribution. On this basis, we introduce varying degrees of \emph{reputation sensitivity}, governed by the parameter \( m \), which modulates the relative weight of reputation in fitness evaluations. Specifically, we consider three levels of sensitivity: \( m = 0 \) (high sensitivity), \( m = 0.5 \) (moderate sensitivity), and \( m = 1 \) (no sensitivity). The resulting evolution of the proportion of cooperators is depicted in Fig.~\ref{fig6}.

Our results consistently demonstrate that the initial distribution of reputation exerts limited influence on the long-term cooperation level. Regardless of the network structure or the degree of reputation sensitivity, the system rapidly converges to a steady state, and the evolution trajectories across different scenarios are remarkably similar. The outcome indicates that the dominant factor governing cooperation dynamics lies in the reputation update and strategy adoption mechanisms themselves, rather than in the initial heterogeneity of individual standing. Moreover, the weak dependence on initial conditions underscores the robustness of the reputation-driven system: cooperators can reliably emerge and stabilize even in environments marked by substantial asymmetries at the outset.

\section{Conclusions}
\label{sec: conclusion}
Our work proposes a reputation-driven game transformation model, aiming to explore the co-evolution mechanism between individual strategy and social status in complex networks. Importantly, we combined a heterogeneous game transformation and a dynamic reputation feedback mechanism and depicted how the behavior choices of individuals in the game process are affected by both their reputation and social interaction. By introducing an adaptive reputation threshold, individuals are dynamically divided into high-reputation and low-reputation groups according to their reputation, and the types of games in which individuals participate are transformed into high-value and low-value games. At the same time, the reputation value of an individual is updated in real-time according to the strategies and his interacting partner in the game, which in turn affects his future game type and strategy evolution.

Based on our results, we have drawn several key conclusions. First, the network structure plays an important regulatory role in the evolution of cooperation. In particular, lattice networks support the long-term coexistence of competing strategies through localized connections, while small-world networks are more susceptible to parameter changes due to their high information dissemination efficiency and sensitivity to strategy evolution. Second, the reputation mechanism promotes the formation of a cooperation dominant state by dynamically adjusting the strategy choices of individuals. Especially in low-value games, cooperation can be maintained even when the temptation factor is high. The threshold change in the reputation mechanism is closely related to the reputation sensitivity of individuals. When the reputation sensitivity is high, the cooperation is more likely to gain support, while when the sensitivity is low, defection is likely to increase. At the same time, the complexity of the network structure also regulates the effectiveness of the reputation mechanism. It is worth noting that although the initial reputation distribution affects the path of cooperation evolution, it has little impact on the final steady state of the system.

In our work, we have studied the phenomenon of reputation-based game transition motivated by real life experiences, but there is still room for further expansion. First, multiple discrete strategies or continuous strategy spaces can be introduced to more realistically characterize the strategy differences of individuals. Second, considering the complexity of the interaction structure in real society, the model can be extended to multiplex networks or temporal networks in the future to explore the impact of multiple relationships and dynamic connections on the evolution of cooperation. In addition, introducing random perturbations in the reputation update mechanism can help simulate information noise and environmental uncertainty in reality, and combine it with more realistic strategy update rules to improve the practical applicability of the model.

\section*{Acknowledgments}
H. Yue, X. Xiong, and M. Feng are supported by grant No. 62206230 funded by the National Natural Science Foundation of China (NSFC), and grant No. CSTB2023NSCQ-MSX0064 funded by the Natural Science Foundation of Chongqing. A. Szolnoki is supported by the National Research, Development and Innovation Office (NKFIH) under Grant No. K142948.               

\bibliographystyle{elsarticle-num-names}
\bibliography{refs}

\end{document}